\documentstyle[12pt,amsfonts,epsfig]{article}
\def \arctanh{\mathop{\rm arctanh}\nolimits}

\newcommand{\beq}{\begin{eqnarray}}
\newcommand{\eeq}{\end{eqnarray}}
\newcommand{\eqn}{\begin{equation}}
\newcommand{\een}{\end{equation}}
\begin{document}
\title{Stability of Kink Defects in a Deformed $O(3)$ Linear Sigma Model}
\author{\dag A. Alonso Izquierdo, \dag M.A.
Gonz\'alez Le\'on \\ and \ddag J. Mateos Guilarte \\ \dag {\small
DEPARTAMENTO DE MATEM\'ATICA APLICADA} \\ \ddag {\small
DEPARTAMENTO DE F\'ISICA} \\ Universidad de Salamanca, SPAIN}
\date{}
\maketitle

\begin{abstract}
We identify the kinks of a deformed $O(3)$ linear Sigma model as
the solutions of a set of first-order systems of equations; the
above model is a generalization of the MSTB model with a
three-component scalar field. Taking into account certain kink
energy sum rules we show that the variety of kinks has the
structure of a moduli space that can be compactified in a fairly
natural way. The generic kinks, however, are unstable and Morse
Theory provides the framework for the analysis of kink stability.
\end{abstract}

\newpage

\section{Introduction}

In Reference \cite{1} we investigated the solitary waves that
arise in a deformed $O(3)$ linear Sigma model. These non-linear
waves appear as kink defects when the system is considered in a
$(1+1)$-dimensional space-time. The research performed in \cite{1}
was based on study of the Hamilton-Jacobi equation of the
mechanical analogue system that follows when time-independent
field configurations are considered. Interest in this model was
explained in our previous work \cite{1}, and the physical meaning
of the kinks was also discussed. Our aim in the present paper is
to gain a better understanding of the nature of such a rich
variety of kinks. We shall focus on three important aspects:

I. In order to clarify the origin of the kink-energy sum rules we
shall develop a treatment \`a la Bogomolny \cite{2} rather than
applying the Hamilton-Jacobi method as in \cite{1}. The system,
which is a generalization of the MSTB model to a three-component
scalar field, is the Bosonic sector of a super-symmetric system:
the interaction energy is derived from a super-potential (in fact,
this property is not exclusive of the three-component case: the
$N$-component analogue model is also the Bosonic sector of a
super-symmetric theory). Thus, the time-independent solutions
satisfy a set of systems of first-order ODE that, of course, is
equivalent to the ODE system obtained in the framework of the
Hamilton-Jacobi paradigm. The bonus of this approach is double:
first, the appearance of the kink energies as several Bogomolny
bounds clarifies the fact that the energy of a generic kink is
equal to the sum of the energies of two or three non-generic kinks
in several ways and, second, the Bogomolny formulation allows us
to prove the stability of one of the kink solutions: the absolute
minimum of the energy in the appropriate topological sector.

II. There is a symmetry group in the system that is generated by
the transformations $\phi_a\rightarrow (-1)^{\delta_{ab}}\phi_b$,
$b=1,2,3$, where $\phi_a$, $a=1,2,3$, are the three components of
the field. This group, $G={\Bbb Z}_2\times{\Bbb Z}_2\times{\Bbb
Z}_2$, acts on the variety of kink solutions, which are classified
into three types; see \cite{1}:

1. \underline{Generic kinks}. There are two three-parametric
families of topological kink trajectories whose three components
are non-null. For obvious reasons, they are usually termed as TK3
kinks. We shall use the term kink orbit, or simply kink, to denote
the kink solution without specifying the spatial dependence. From
this point of view, TK3 kinks are two two-parametric families of
kink orbits and two three-parametric families of kink trajectories
or solitary waves.

2. \underline{Enveloping kinks}. There are four one-parametric
families of non-topological kink orbits that also have three
non-null components. NTK3 kinks live on one ellipsoid in the
${\Bbb R}^3$ internal space that encloses all the other kinks.

3. \underline{Embedded kinks}. All the solitary waves of the $N=2$
MSTB model appear embedded twice in the kink manifold of the $N=3$
system. Thus, these kinks have at most two non-null components.
There are two classes: a) Two-component non-topological and
topological kinks. The NTK2 kink orbits form two one-parametric
families in the $\phi_3=0$ plane and another two one-parametric
families in the $\phi_2=0$ plane. TK2 kinks appear at a limit of
these two families that we shall explain later. b) One-component
topological kinks. There exists one topological TK1 kink and one
anti-kink both of which live on the $\phi_1$-axis.

We briefly describe the action of $G$ on a given kink, starting in
the opposite order:
\begin{itemize}
\item One-component topological kinks and anti-kinks are fixed
points of the action of the elements $\phi_2\rightarrow -\phi_2$,
$\phi_3\rightarrow -\phi_3$. The transformation $\phi_1\rightarrow
-\phi_1$, however, sends a TK1 kink to its anti-kink, and
vice-versa.
\item The NTK2 families are invariant under the action of either
$\phi_3\rightarrow -\phi_3$ or $\phi_2\rightarrow -\phi_2$.
$\phi_1\rightarrow -\phi_1$ exchanges the families living in the
same plane. Finally, $\phi_2\rightarrow -\phi_2$ sends one NTK2
kink of the $\phi_3=0$ plane to another in the same family, all of
them invariant under $\phi_3\rightarrow -\phi_3$. We find the same
situation with the NTK2 family in the $\phi_2=0$ plane with
respect to $\phi_3\rightarrow -\phi_3$ and $\phi_2\rightarrow
-\phi_2$, respectively.
\item The four NTK3 families are interchanged through the action
of $\phi_1\rightarrow -\phi_1$ and $\phi_3\rightarrow -\phi_3$.
$\phi_2\rightarrow -\phi_2$, however, relates two NTK3 kinks in
the same family.
\item The two TK3 families are related by the $\phi_1\rightarrow -\phi_1$
transformation. $\phi_2\rightarrow -\phi_2$ and $\phi_3\rightarrow
-\phi_3$ link two TK3 kinks in the same family.
\end{itemize}

The moduli space of kinks is the quotient of the variety of kinks
by the action of the $G$ group. Restriction to the sub-variety of
generic kinks leads to a good structure for the kink moduli
sub-space because there are no fixed points. We shall show (see
also \cite{1}) that the kink solutions are obtained using Jacobi
elliptic coordinates in the internal ${\Bbb R}^3$ space. The
sub-space of TK3 kink orbits in elliptic coordinates is
parametrized by two real integration constants. Thus, the TK3
moduli sub-space is the open ${\Bbb R}^2$ plane. Because the
change from Cartesian to elliptic coordinates is generically a
$2^3$-to-$1$ map, a point in ${\cal M}_{\rm TK3}={\Bbb R}^2$
corresponds to eight TK3 kinks, which precisely form an orbit -in
the sense of group action- of the $G$ group.

There are identities between the energy of a generic kink and the
sum of one embedded kink and one enveloping kink, or the sum of
several embedded kinks, in several ways. The combinations of the
embedded and the enveloping kinks are singular in the sense that
they are fixed points of some sub-group of $G$. Fortunately, one
can see that the singular configurations are a limiting case of
TK3 kinks that appear in the \lq\lq boundary" of the TK3 moduli
sub-space. This situation calls for a compactification of ${\cal
M}_{\rm TK3}$ and we shall describe how it is possible to include
the whole variety of kinks in a compact moduli space.

We thus find analogies with the Mumford-Deligne compactification
of the moduli space of Riemann surfaces of genus $g$, \cite{4}.
Also, the kink moduli space is a similar structure to the moduli
space of instantons in pure gauge theory, \cite{5}, or of BPS
monopoles in Yang-Mills-Higgs systems, \cite{6}: the discrete
group $G$ replaces the infinite dimensional diffeomorphism group
in the case of surfaces, or the gauge group in the other cases.
The orbit of every point in the kink moduli space is a discrete
set; the main difference with respect to the gauge theoretical
analogues, however, is that in the system under consideration
there are no solutions with several kinks and therefore the moduli
space only encompasses the sectors of topological charges $Q_T=\pm
1$.

III. There is another difference with the gauge theory solitons
mentioned above: both the instantons and the BPS magnetic
monopoles are absolute minima of either the Euclidean action or
the energy, and are thus completely stable. The TK3 kinks,
however, are not absolute minima of the energy; in fact, we will
show that they are unstable. Study of the stability of the
different types of kinks is the main concern of this paper. We
shall analyze the stability problem, which is crucial in order to
envisage the nature of the quantum states built around the
classical kinks in three stages:
\begin{itemize}
\item The eight super-potentials that determine eight systems of
first-order equations cannot be differentiated at two \lq\lq
focal" lines: an ellipse in the $\phi_3=0$ plane and a hyperbola
in the orthogonal $\phi_2=0$ plane. Kink solutions crossing by any
of these curves fail to be absolute minima of the energy in their
topological sectors: these solutions are obtained by gluing the
solutions of at least two different systems of first-order
equations. We shall also show that there are Jacobi fields along
these kink orbits with zeroes in the vacuum points and the
intersection points with the focal lines. According to one Jacobi
theorem, see \cite{16}, these kinks are not local minima of the
energy and are therefore unstable.
\item The direct approach to studying kink stability requires an
analysis of the spectrum of the second variation, or Hessian
functional. The explicit expression of the Hessian functional is
only available for the non-generic topological kinks. In the
Schrodinger operator that defines the Hessian quadratic form, the
potential well depends on the kink solution; the kink solutions,
however, cannot be written in terms of elementary transcendental
functions except in the non-generic cases mentioned above. In
fact, besides the Jacobi fields, the spectral problem can only be
fully solved for the TK1 kink, and, partially, for the
TK2$\sigma_3$ kink. Nevertheless, the information obtained this
way fits in with the kink-energy sum rules and with the fact that
these non-generic kinks live at the boundary of the TK3 kink
moduli space perfectly well.

\item  Finally, we shall construct the Morse theory for the
configuration space, choosing energy as the Morse functional.
Knowledge of the Jacobi fields allows an immediate application of
the Morse index theorem, as in Reference \cite{8}, to identify the
Morse index - the dimension of the negative eigen-space of the
Hessian - of a kink trajectory with the number of crossing points
through the focal ellipse and hyperbola: this count measures the
degree of kink instability. Moreover, we can easily proceed and
read the homology of the configuration space of the system from
the critical point structure of the Hessian, following the pattern
developed in \cite{9} for the MSTB model. This procedure affords a
direct connection between the existence and fate of topological
defects and the topology of the configuration space, which, in
turn, reveals many features of the quantum states built from these
classical extended objects.
\end{itemize}

The paper is organized in three Sections. There are three
sub-sections in Section \S 2. \S 2.1 introduces the model and
proposes that the solitary waves satisfy a set of systems of
first-order ordinary differential equations. \S 2.2 solves the
system by passing to Jacobi elliptic coordinates. \S 2.3 discusses
the existence of the (several) super-potentials that allow the
reduction to the first-order ODE system, describes the kink
solutions, and establishes the stability of the absolute minimum
of the energy functional. Section \S 3 is divided into three
sub-sections. \S 3.1 states the kink-energy sum rules. In \S 3.2
the singular kinks are considered as the limit of generic kinks
when some of the integration constants that characterize the
variety of solutions tend to infinity. In \S 3.3 we briefly
discuss the results of the two previous sub Sections by showing
the compact kink moduli space. The theme of Section \S 4 is the
stability of kinks. In \S 4.1 we give explicit expressions for the
Jacobi fields along the TK3 and NTK trajectories. The zeroes of
these fields occur when the kink trajectories cross two focal
lines, one an ellipse and the other one an hyperbola. Sub-section
\S 4.2 is devoted to solving the spectral problem for the
second-order fluctuation operator around the non-generic
topological kinks. In \S 4.3, the topological implications of
stability are unveiled through the elaboration of a Morse theory
of kinks. Finally, some reflections on kink quantization are
offered in Section \S 5.

\section{Bogomolny equations}
Besides the second-order evolution equations, many soliton-like
solutions satisfy first-order PD or OD equations in relativistic
field theories with relevance in fundamental physics and
Cosmology. This interesting discovery is due to a generalization
by Bogomolny, \cite{2}, of the self-duality equations of Euclidean
gauge theory, \cite{5}, to other space-time dimensions and other
systems. Bogomolny's idea has deep topological roots and leads to
very rich spaces of (BPS) states in quantum field theory with
extended super-symmetry, see \cite{10}. The purpose of this
Section is to show how the kinks discovered in Reference \cite{1}
in a deformed $O(3)$ linear Sigma model also enter this framework.

\subsection{The model: kink defects and first-order equations}
We start by briefly describing the model and characterizing the
solitary waves as solutions of a system of first-order ODE.
Defining non-dimensional space-time coordinates, fields and
physical parameters as in \cite{1}, the dynamics of the model is
determined by the action functional
\begin{equation}
S=\frac{m^2}{\lambda^2} \int d^2 x \left\{ \frac{1}{2}
\partial_\mu \vec{\phi}\cdot \partial^\mu\vec{\phi}-\frac{1}{2} \left( \vec{\phi}\cdot
\vec{\phi}-1\right)^2 -\sum_{a=1}^3\frac{1}{2}
\sigma_a^2\phi_a^2\right\}\label{1}
\end{equation}
We also use all the conventions in \cite{1} and focus on the
maximally asymmetric case, $\sigma_1^2=0<\sigma_2^2<\sigma_3^2<1$,
in the range of the manifold parameters $\sigma_2^2$ and
$\sigma_3^2$, where the kink manifold is richest.

The mechanical system associated with the search for the solitary
wave solutions of the model is completely integrable in the sense
of Liouville and all the kink solutions can be found analytically.
Here we shall not pursue this line of research; it has been fully
developed in \cite{1}. Instead, we assume that there exists a
super-potential for the system; i.e., a function
$W(\vec{\phi}):{\Bbb R}^3\rightarrow {\Bbb R}$ in the internal
space such that:
\begin{equation}
\sum_{a=1}^3\frac{\partial W}{\partial \phi_a}\frac{\partial
W}{\partial \phi_a}=\left (\vec{\phi}\cdot\vec{\phi}-1\right
)^2+\sum_{a=1}^3\sigma_a^2\phi_a^2\label{12}
\end{equation}
If this is the case, it is possible to write the energy for static
configurations ($\phi_a=\phi_a(x)$) \`a la Bogomolny, see
\cite{2}:
\begin{eqnarray}
E[\vec{\phi}]&=& \frac{m^3}{\lambda^2\sqrt{2}}\int dx \left\{
\frac{1}{2} \frac{d\vec{\phi}}{dx}\cdot
\frac{d\vec{\phi}}{dx}+\frac{1}{2} \left( \vec{\phi} \cdot
\vec{\phi}-1\right)^2+\frac{1}{2}\sum_{a=1}^3\sigma_a^2
\phi_a^2\right\}=\nonumber\\ &=&
\frac{m^3}{2\lambda^2\sqrt{2}}\left [\int dx \sum_{a=1}^3\left
(\frac{d\phi_a}{dx}- \frac{\partial W}{\partial\phi_a}\right
)\left (\frac{d\phi_a}{dx}- \frac{\partial
W}{\partial\phi_a}\right )+ 2\int
\sum_{a=1}^3d\phi_a\frac{\partial W}{\partial \phi_a}\right
]\label{13}
\end{eqnarray}
Configurations that satisfy the system of first-order equations:
\begin{equation}
\frac{d\phi_a}{dx}=\frac{\partial W}{\partial\phi_a},\label{14}
\end{equation}
do not contribute to the first term in $E[\vec{\phi}]$ and are
also solutions of the second-order equations of the model. The
solutions of (\ref{14}) are kinks of the system if they also
comply with the finite energy conditions:
\begin{equation}
\lim_{x\to \pm \infty}\frac{d\phi_a}{dx}=0\  ,\qquad \lim_{x\to
\pm \infty}\phi_a(x)=v_a,\  \forall a=1,2,3 .\label{15}
\end{equation}
where $\vec{v}\equiv (v_1,v_2,v_3)$ is a vector that belongs to
the vacuum manifold ${\cal V}$. ${\cal V}$ is formed by two
vectors $\vec{v}^\pm \equiv (\pm 1,0,0)$ and the existence of two
points in ${\cal V}$ classifies the kink solutions into
topological and non-topological, see \cite{1} for details.

Every kink orbit traces out a path in the internal ${\Bbb R}^3$
space. There are two possibilities:
\begin{itemize}
\item The super-potential $W(\vec{\phi})$ is differentiable along
the kink path. Thus, the second term in (\ref{13}) is the integral
of an exact differential and Stoke's theorem tells us that it
depends on the difference of the values of $W$ at the path
endpoints. The energy is a topological bound and the kinks of this
type are absolute minima of $E[\vec{\phi}]$ and stable against
small fluctuations.
\item There is a discrete set of points along the kink path where
$W(\vec{\phi})$ is not differentiable. The second term in
(\ref{13}) is only differentiable piece-wise and $E[\vec{\phi}]$
also depends on the values of $W(\vec{\phi})$ at those points of
non-differentiability; for this type of kinks, $E[\vec{\phi}]$ is
not a topological bound. We cannot say strictly that these kinks
are solutions of (\ref{14}) because the points of
non-differentiability of $W(\vec{\phi})$ are turning points of the
the kink orbit. Moreover, in sub-Section \S{2.3} we shall show
that there are eight possible choices of $W$. If $\vec{\phi}(x_0)$
is a point where $W$ is non-differentiable, a kink is a solution
of (\ref{14}) for $x\in (x_0-\epsilon,x_0)$, with a given choice
of $W$; the same kink solves (\ref{14}) with another choice of $W$
for $x\in (x_0,x_0+\epsilon)$. These kinks are solutions of the
second-order equations but are not absolute minima of
$E[\vec{\phi}]$. We shall see that not only the non-topological
kinks belong to this class, but also many others that live in the
topological sectors.
\end{itemize}

\subsection{Jacobi elliptic coordinates}
We now introduce Jacobi elliptic coordinates in the internal
${\Bbb R}^3$ space, see \cite{1}. Defining ${\bar
\sigma_a}^2=1-\sigma_a^2$, we shall denote by ${\bf P}_3(\infty)$
the interior of the infinite parallelepiped ${\bf {\bar
P}}_3(\infty)=\partial {\bf {\bar P}}_3(\infty )\bigsqcup {\bf
P}_3(\infty )$:
\begin{equation}
-\infty < \lambda_1 \leq {\bar \sigma}_3^2 \leq \lambda_2 \leq
{\bar \sigma}_2^2 \leq \lambda_3 \leq 1 , \label{16}
\end{equation}

Thus the map $\rho : \vec{\phi} \longrightarrow \vec{\lambda}$,
from ${\Bbb R}^3$ to ${\bf P}_3(\infty)$, given by the change of
coordinates:
\begin{eqnarray}
\phi_1^2&=&\frac{(1-
\lambda_1)(1-\lambda_2)(1-\lambda_3)}{\sigma_2^2\sigma_3^2},\hspace{1.0cm}
\phi_2^2=
\frac{(\bar{\sigma}_2^2-\lambda_1)(\bar{\sigma}_2^2-\lambda_2)
(\bar{\sigma}_2^2-\lambda_3)}{-\sigma_2^2(\sigma_3^2-\sigma_2^2)}\nonumber\\
\phi_3^2&=&
\frac{(\bar{\sigma}_3^2-\lambda_1)(\bar{\sigma}_3^2-\lambda_2)
(\bar{\sigma}_3^2-\lambda_3)}{\sigma_3^2(\sigma_3^2-\sigma_2^2)}\label{17}
\end{eqnarray}
sends eight points of ${\Bbb R}^3$ to one point in ${\bf
P}_3(\infty )$ and induces a Riemannian metric in ${\bf
P}_3(\infty )$:
$g_{aa}(\vec{\lambda})=\frac{-f_a(\vec{\lambda})}{4 A(\lambda_a)},
a=1,2,3,\quad g_{ab}=0, \forall a\neq b \quad$, where:
$A(\lambda_a)=(\lambda_a-1)(\lambda_a-\bar{\sigma}_2^2)(\lambda_a-\bar{\sigma}_3^2)$,
and $f_a(\vec{\lambda})=\displaystyle \prod_{b=1\atop a\neq
b}^3(\lambda_a-\lambda_b)$.

The potential energy, $V(\vec{\phi})=\frac{1}{2}\left(
\vec{\phi}\cdot \vec{\phi}-1\right)^2+\frac{1}{2} \sigma_2^2
\phi_2^2+\frac{1}{2} \sigma_3^2 \phi_3^2$, in elliptic coordinates
reads:
\begin{equation}
V(\vec{\lambda})=\frac{1}{2}\left(
\frac{\lambda_1^2(\lambda_1-\bar{\sigma}_2^2)
(\lambda_1-\bar{\sigma}_3^2)}{(\lambda_1-\lambda_2)(\lambda_1-\lambda_3)}+
\frac{\lambda_2^2(\lambda_2-\bar{\sigma}_2^2)
(\lambda_2-\bar{\sigma}_3^2)}{(\lambda_2-\lambda_1)(\lambda_2-\lambda_3)}+
\frac{\lambda_3^2(\lambda_3-\bar{\sigma}_2^2)(\lambda_3-\bar{\sigma}_3^2)}{(\lambda_3-\lambda_1)(\lambda_3-\lambda_2)}\right)\label{18}
\end{equation}

The crucial question is the following: is there a function
$W(\vec{\lambda})$ such that the potential $V$ can be written in
the form
\begin{equation}
2\, V(\vec{\lambda})= g^{-1}_{11}\frac{\partial W}{\partial
\lambda_1}\frac{\partial W}{\partial
\lambda_1}+g^{-1}_{22}\frac{\partial W}{\partial
\lambda_2}\frac{\partial W}{\partial \lambda_2}+
g^{-1}_{33}\frac{\partial W}{\partial \lambda_3}\frac{\partial
W}{\partial \lambda_3}   \  ? \label{eqq}
\end{equation}
The answer is affirmative if the super-potential
$W(\vec{\lambda})$ is a solution of the PDE:
\begin{equation}
\sum_{a=1}^3\frac{-4 (\lambda_a
-1)(\lambda_a-\bar{\sigma}_2^2)(\lambda_a-
\bar{\sigma}_3^2)}{f_a(\vec{\lambda})} \left( \frac{\partial
W}{\partial \lambda_a}\right)^2= \sum_{a=1}^3
\frac{\lambda_a^2(\lambda_a-\bar{\sigma}_2^2)(\lambda_a-\bar{\sigma}_3^2)}{f_a(\vec{\lambda})}\label{19}
\end{equation}

Note that (\ref{19}) is no more than the Hamilton-Jacobi equation,
formula (20) in \cite{1}, for the Hamilton characteristic function
$W(\vec{\lambda})$, with no explicit dependence on $x$.

Searching for solutions of the form
$W(\vec{\lambda})=W_1(\lambda_1)+W_2(\lambda_2)+W_3(\lambda_3)$,
(\ref{19}) becomes three separate ODE:
\begin{equation}
\left( \frac{d W_a}{d \lambda_a}\right)^2=\frac{\lambda_a^2}{4
(1-\lambda_a)},\quad \forall a=1,2,3\label{20}
\end{equation}
the ordinary differential equations into which the HJ equation
degenerates when all the separation constants are at zero, see
formula (29) for $N=3$ in \cite{1}.

The integration of (\ref{20}) is elementary:
\[
W_a^{(\alpha_a )}=(-1)^{\alpha_a} \int \frac{\lambda_a
d\lambda_a}{2\sqrt{1-\lambda_a}}=-(-1)^{\alpha_a} \frac{1}{3}
(\lambda_a+2) \sqrt{1-\lambda_a},\hspace{1cm} \alpha_a=0,1 ,
\]
if we set the integration constant to be zero. There is also a
sign ambiguity fixed by the choice of $\alpha_a$. Thus, there are
eight solutions of the PDE (\ref{19}):
\begin{equation}
W^{(\alpha_1, \alpha_2, \alpha_3 )}(\vec{\lambda}) = \sum_{a=1}^3
 (-1)^{\alpha_a}\frac{1}{3}  (\lambda_a+2) \sqrt{1-\lambda_a} \label{21}
\end{equation}
The generalization of equation (\ref{eqq}) to the $N$-component
scalar field case is solved using the same antsatz:
$W=\sum_{a=1}^N W_a$ leads us to the solution:
$W^{(\alpha_1,\dots,\alpha_N)}(\vec{\lambda})=\sum_{a=1}^N
(-1)^{\alpha_a} \frac{1}{3} (\lambda_a+2) \sqrt{1-\lambda_a} $.

The energy for static configurations is written in elliptic
coordinates as:
\begin{equation}
E[\vec{\lambda}]=\frac{m^3}{2\lambda^2\sqrt{2}} \int dx
\sum_{a,b=1}^3 \left\{ g_{ab}(\vec{\lambda})
\frac{d\lambda_a}{dx}\cdot
\frac{d\lambda_b}{dx}+g_{ab}^{-1}(\vec{\lambda})\frac{\partial
W^{(\alpha_1,\alpha_2,\alpha_3)}}{\partial\lambda_a}\frac{\partial
W^{(\alpha_1,\alpha_2,\alpha_3)}}{\partial
\lambda_b}\right\}\label{22}
\end{equation}
The possibility of writing the potential energy as a \lq\lq
square" in this way means that the (1+1)-dimensional field theory
system admits a $N=1$ super-symmetric extension. Here we shall not
discuss the super-symmetric system; instead, we focus on the fact
that the energy can be written \`a la Bogomolny \cite{2}:
\begin{eqnarray}
E[\vec{\lambda}]&=&\frac{m^3}{2\lambda^2\sqrt{2}}\left [ \int dx
\sum_{a,b=1 }^3 g_{ab}(\vec{\lambda}) \left [
\frac{d\lambda_a}{dx} + \sum_{c=1}^3
g_{ac}^{-1}(\vec{\lambda})\frac{\partial
W^{(\alpha_1,\alpha_2,\alpha_3 )}}{\partial \lambda_c}\right ]
\left [ \frac{d\lambda_b}{dx} + \sum_{d=1}^3
g_{bd}^{-1}(\vec{\lambda})\frac{\partial
W^{(\alpha_1,\alpha_2,\alpha_3 )}}{\partial \lambda_d}\right
]\right ] \nonumber\\ &+&\frac{m^3}{\lambda^2\sqrt{2}}\left [\left
| \int dx \sum_{a=1}^3 \frac{\partial
W^{(\alpha_1,\alpha_2,\alpha_3
)}}{\partial\lambda_a}[\vec{\lambda}] \frac{d\lambda_a}{dx}\right
| \right ] \label{23}
\end{eqnarray}
The first integral in (\ref{23}) gives a semi-definite positive
contribution and there is a bound to the energy of a kink that
satisfies the inequality: $E[\vec{\lambda}]\geq
\frac{m^3}{\lambda^2\sqrt{2}} \displaystyle \left| \int_P
\sum_{a=1}^3\frac{\partial W^{(\alpha_1,\alpha_2,\alpha_3
)}}{\partial\lambda_a}[\vec{\lambda}] d\lambda_a\right| $. The
Bogomolny bound is saturated -the inequality becomes equality- if
the \lq\lq first-order" equations
\begin{eqnarray}
\frac{d\lambda_1}{dx}&=& (-1)^{\alpha_1} 2 \frac{\lambda_1
(\lambda_1-\bar{\sigma}_2^2)
(\lambda_1-\bar{\sigma}_3^2)}{(\lambda_1-\lambda_2)(\lambda_1-\lambda_3)}\cdot
\sqrt{1-\lambda_1}\nonumber\\ \frac{d\lambda_2}{dx}&=&
(-1)^{\alpha_2} 2 \frac{\lambda_2 (\lambda_2-\bar{\sigma}_2^2)
(\lambda_2-\bar{\sigma}_3^2)}{(\lambda_2-\lambda_1)(\lambda_2-\lambda_3)}\cdot
\sqrt{1-\lambda_2}\label{24}\\ \frac{d\lambda_3}{dx}&=&
(-1)^{\alpha_3} 2 \frac{\lambda_3 (\lambda_3-\bar{\sigma}_2^2)
(\lambda_3-\bar{\sigma}_3^2)}{(\lambda_3-\lambda_1)(\lambda_3-\lambda_2)}\cdot
\sqrt{1-\lambda_3}\nonumber
\end{eqnarray}
are satisfied. We see that the existence of eight superpotentials
is related to the choice of signs in the right members of the ODE
system (\ref{24}). Alternatively, we could fix the superpotential,
e.g. by setting $\alpha_1=\alpha_2=\alpha_3=0$, and allow for an
independent choice of signs in each equation in (\ref{24}) or in
the Bogomolny splitting (\ref{23}). In any case, (\ref{24})
constitutes eight different systems of three ordinary differential
equations.

We write the system (\ref{24}) in the form:
\begin{equation}
\frac{d\lambda_a}{(-1)^{\alpha_a}2\lambda_a(\lambda_a-{\bar\sigma}_2^2)
(\lambda_a-{\bar\sigma}_3^2)\sqrt{1-\lambda_a}}=\frac{dx}{f_a(\vec{\lambda})}
;\qquad a=1,2,3 \label{25}
\end{equation}
The sum of the three equations in (\ref{25}) reads:
\begin{equation}
\displaystyle\sum _{a=1}^3
\frac{d\lambda_a}{(-1)^{\alpha_a}2\lambda_a(\lambda_a-{\bar\sigma}_2^2)
(\lambda_a-{\bar\sigma}_3^2)\sqrt{1-\lambda_a}}=0 \label{26}
\end{equation}
We can also re-organize the equation (\ref{25}) multiplying both
members by $\lambda_a-\lambda_b$, $a\neq b$. Thus, this equivalent
form of  the ODE system (\ref{25}) contains six equations that can
be added to obtain:
\begin{equation}
\displaystyle\sum _{a=1}^3 \frac{\lambda_a
d\lambda_a}{(-1)^{\alpha_a}2\lambda_a(\lambda_a-{\bar\sigma}_2^2)
(\lambda_a-{\bar\sigma}_3^2)\sqrt{1-\lambda_a}}=0 , \label{27}
\end{equation}
through the use of the three equations in (\ref{25}). Integration
of (\ref{26}) and (\ref{27}) provides all the separatrix orbits of
the mechanical analogue system obtained by means of the
Hamilton-Jacobi principle applied for zero particle energy and
zero separation constants: compare with equations (31) and (32)
for $N=3$ in Reference \cite{1}. Therefore, all the orbits found
there are the kink solutions of the first-order system (\ref{24}).
The kink form factors -the kink trajectories in the dynamical
system terminology- are obtained after multiplication of the three
equations in (\ref{25}) by $f_a(\vec{\lambda})$, addition of the
three resulting identities, and use of (\ref{26}) and (\ref{27}):
\begin{equation}
\displaystyle\sum _{a=1}^3 \frac{\lambda_a^2
d\lambda_a}{(-1)^{\alpha_a}2\lambda_a(\lambda_a-{\bar\sigma}_2^2)
(\lambda_a-{\bar\sigma}_3^2)\sqrt{1-\lambda_a}}=dx , \label{29}
\end{equation}
Note that the boundary conditions (\ref{15}) in elliptic
coordinates read:
\begin{equation}
\lim_{x\to \pm \infty}\frac{d\lambda_a}{dx}=0\  ,\qquad \lim_{x\to
\pm \infty}\lambda_1(x)=0\ , \qquad \lim_{x\to \pm
\infty}\lambda_2(x)={\bar\sigma}_3^2\ ,\qquad \lim_{x\to \pm
\infty}\lambda_3(x)={\bar\sigma}_2^2\ .\label{28}
\end{equation}
Moreover, by squaring the first equation in (\ref{24}) and
defining the generalized momentum
$\pi_1=g_{11}(\vec{\lambda})\frac{d\lambda_1}{dx}$, we obtain:
\begin{equation}
\frac{1}{2}\pi_1^2+\frac{1}{8}\frac{\lambda_1^2}{\lambda_1-1}=0\label{36}
\end{equation}
Equation (\ref{36}) describes the motion of a particle with zero
energy moving under the influence of a potential
\begin{eqnarray}
{\cal U}(\lambda_1)&=&
\frac{1}{4}\frac{\lambda_1^2}{\lambda_1-1},\hspace{1cm}
 -\infty<\lambda_1\leq {\bar \sigma}_3^2\nonumber\\&=&
 \infty,\hspace{2,2cm}
  {\bar \sigma}_3^2<\lambda_1<\infty\nonumber
\end{eqnarray}
${\cal U}(\lambda_1)$ has a maximum at $\lambda_1=0$ and goes to
$-\infty$ when $\lambda_1$ tends to $-\infty$; therefore, bounded
motion occurs only in the $\lambda_1\in [0,{\bar \sigma}_3^2]$
interval. Together with the boundary conditions (\ref{28}), this
means that the kink configurations lie in the finite
parallelepiped ${\bf{\bar P}}_3(0)$:
$0\leq\lambda_1\leq{\bar\sigma}_3^2\leq\lambda_2\leq{\bar\sigma}_2^2\leq\lambda_3\leq1$.

\subsection{Kink solutions and Kink energies} In this
section we shall analyze the solutions of the first order
equations (\ref{24}). The change in coordinates from Cartesian to
elliptic is singular at each face in ${\bf{\bar P}}_3(0)$, except
at the dynamical frontier $\lambda_1=0$. We shall thus deal
carefully with the limiting behaviour of the system in $\partial
{\bf{\bar P}}_3(0)$. There is a dimensional reduction of the
system at these points and we expect a different kind of behaviour
with respect to the regular points of the interior of ${\bf{\bar
P}}_3(0)$. Computation of the energies of the kink solutions can
be performed by noticing that the solutions of the first-order
equations (\ref{24}) saturate the Bogomolny bound:
$E[\vec{\lambda}_K]=\left|\int_K
dW^{(\alpha_1,\alpha_2,\alpha_3)}\right| $

\subsubsection{Generic Kinks.} Integration of (\ref{26}) and
(\ref{27}) provides the orbits that the generic kinks trace in the
internal space ${\Bbb R}^3$:
\begin{equation}
\displaystyle\sum _{a=1}^3\frac{(-1)^{\alpha_a}}{2}\int
\frac{d\lambda_a}{\lambda_a(\lambda_a-{\bar\sigma}_2^2)
(\lambda_a-{\bar\sigma}_3^2)\sqrt{1-\lambda_a}}=\gamma_2
\label{57}
\end{equation}
\begin{equation}
\displaystyle\sum _{a=1}^3\frac{(-1)^{\alpha_a}}{2}\int
\frac{\lambda_a d\lambda_a}{\lambda_a(\lambda_a-{\bar\sigma}_2^2)
(\lambda_a-{\bar\sigma}_3^2)\sqrt{1-\lambda_a}}=\gamma_3
\label{58},
\end{equation}
where $\gamma_2$ and $\gamma_3$ are real integration constants.
The kink form factor is obtained from integration of the equation
(\ref{29}),
\begin{equation}
\displaystyle\sum _{a=1}^3\frac{(-1)^{\alpha_a}}{2}\int
\frac{\lambda_a^2d\lambda_a}{\lambda_a(\lambda_a-{\bar\sigma}_2^2)
(\lambda_a-{\bar\sigma}_3^2)\sqrt{1-\lambda_a}}=\gamma_1+x
\label{59},
\end{equation}
and depends on a third integration constant $\gamma_1$. Thus, the
solitary kink waves of our system are composed of two ingredients:
the orbit and the form factor.

The explicit integrations in (\ref{57}) and (\ref{58}) are
performed in Reference \cite{1}, where they are written in compact
form in formulas (35) and (36). All the ${\rm TK3}$ kink orbits
found there are described in sub-Section \S 3.2 of \cite{1} by
means of a numerical algorithm implemented in Mathematica (here we
reproduce Figure 1 of a generic TK3 in both Cartesian and elliptic
space). We refer the reader to that paper for information about
the ${\rm TK3}$ curves.

Instead, we now focus on computing the energy of a ${\rm TK3}$
generic kink as a Bogomolny bound: generic ${\rm TK3}$ orbits grow
from several steps, according to the different choices of signs in
equations (\ref{26})-(\ref{27}). Each piece of any ${\rm TK3}$
orbit is a solution of the equations (\ref{26})-(\ref{27}), such
that on these curves the ${\rm TK3}$ kink also complies with
(\ref{29}). The signs in each stage, i.e., the values of
$\alpha_a$ in the equations (\ref{26})-(\ref{27})-(\ref{29}), must
be chosen according to the sense of change in the elliptic
variables along the corresponding piece of orbit.

\begin{figure}[htbp]
\begin{center}
\epsfig{file=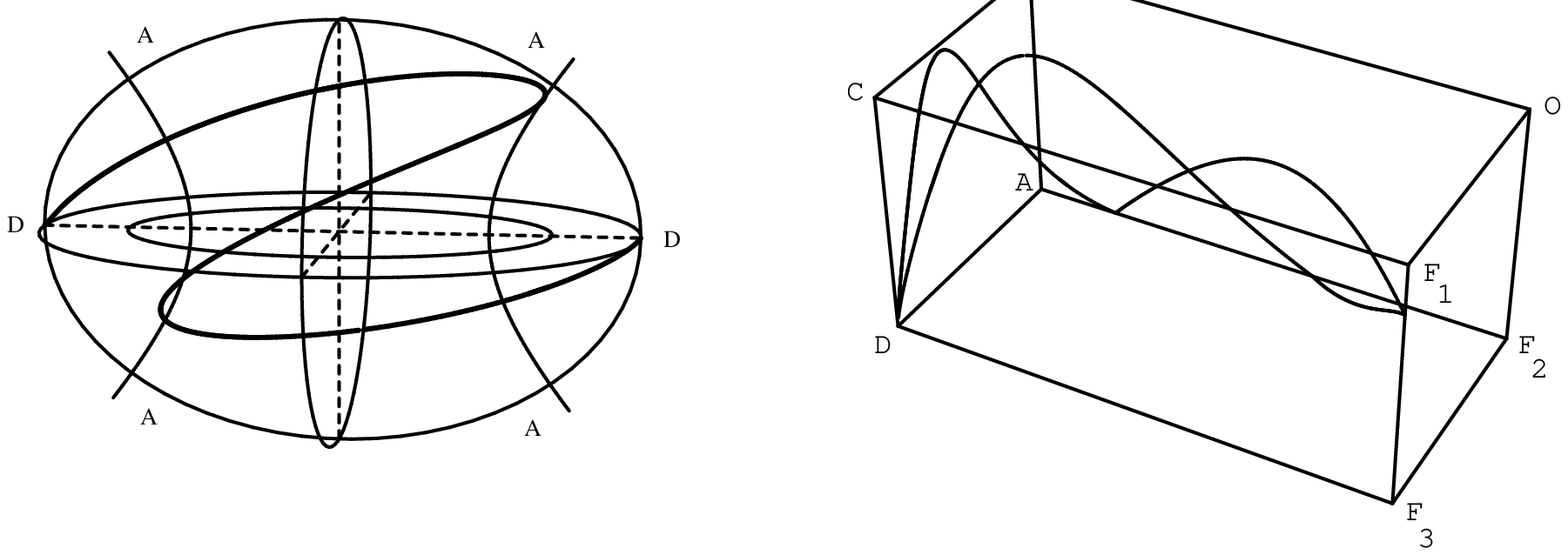,height=4.5cm}
\end{center}
\caption[Figura]{\small Orbit of a TK3 in ${\Bbb R}^3$ and
$\bar{\bf P}_3(0)$.}
\end{figure}
The behaviour of a generic TK3 solution in $\bar{\bf P}_3(0)$ is
as follows:

\begin{itemize}
\item 1. Starting from the point D, with coordinates in $\bar{\bf
P}_3(0)$: D$\equiv(0,\bar{\sigma}_3^2,\bar{\sigma}_2^2)$ at $x\to
-\infty$, the first step travels in ${\bf P}_3(0)$ to a point that
we shall denote as P$_1$, the intersection of the kink with the
face $\lambda_3=1$ in $\partial{\bf P}_3(0)$, attained for some
$x=x_1$ (note that the invariance under translations in $x$, i.e.,
the arbitrariness of fixing $\gamma_1$, makes the exact value of
$x_1$ arbitrary too). The coordinates of P$_1$ will be P$_1\equiv
(\lambda_1^{{\rm P}_1},\lambda_2^{{\rm P}_1},1)$, and hence for
$x\in (-\infty,x_1]$ we have that $\lambda_1,\lambda_2$ and
$\lambda_3$ increase. Thus, $\frac{d\lambda_1}{dx},
\frac{d\lambda_2}{dx}$ and $\frac{d\lambda_3}{dx}$ are positive
for this range and we may conclude that from D to P$_1$ the
first-order equations (\ref{24}) are satisfied if we consider
$\alpha_1=\alpha_2=\alpha_3=0$ in the choice of the
super-potential $W^{(\alpha_1,\alpha_2,\alpha_3)}$.

\item 2. In the second step, the TK3 returns to the interior of the
parallelepiped, ${\bf P}_3(0)$, finishing at point P$_2$, the
intersection with the edge AF$_2$. The coordinates of P$_2$ are
P$_2\equiv (\lambda_1^{{\rm
P}_2},\bar{\sigma}_2^2,\bar{\sigma}_2^2)$ and this is reached for
a value $x=x_2>x_1$. In a similar way to the analysis performed
for the first step, we have that for $x\in(x_1,x_2]$, i.e. from
P$_1$ to P$_2$, equations (\ref{24}) are satisfied only if we
consider the super-potential $W^{(0,0,1)}$.

\item 3. In the third step, the kink travels from the AF$_2$ edge
to the F$_1$F$_3$ one, which is reached at point
P$_3\equiv(\bar{\sigma}_3^2,\bar{\sigma}_3^2,\lambda_3^{{\rm
P}_3})$ when $x=x_3$. The behaviour of the variables impose the
super-potential $W^{(0,1,0)}$ in equations (\ref{24}).

\item 4. From P$_3$ to P$_4$, again in the face $\lambda_3=1$, we
obtain the super-potential $W^{(1,0,0)}$.

\item 5. In step 5, we find a similar situation to step 2,
but the variable $\lambda_1$ decreases, and hence we need
$W^{(1,0,1)}$ from P$_4$ to P$_5$.

\item 6. Denoting the third crossing of the $\lambda_3=1$
face of $\partial {\bf P}_3(0)$ as P$_6$, we have that from P$_5$
to P$_6$ the correct super-potential to be considered is
$W^{(1,1,0)}$.

\item 7. Finally, from P$_6$ to point D we reproduce the first
step in the opposite order, for the interval $x\in (x_6,\infty)$.
\end{itemize}

With all these considerations, the energy can be computed, step by
step, as follows:
\begin{eqnarray*}
\frac{\lambda^2\sqrt{2}}{m^3}\, E({\rm TK3})&=& -\int_{{\rm
TK3}}dW^{(\alpha_1,\alpha_2,\alpha_3)}= -\int_{{\rm D}}^{{\rm
P}_1} dW^{(0,0,0)}-\int_{{\rm P}_1}^{{\rm P}_2}
dW^{(0,0,1)}-\int_{{\rm P}_2}^{{\rm P}_3} dW^{(0,1,0)}\\ &&
-\int_{{\rm P}_3}^{{\rm P}_4} dW^{(1,0,0)}-\int_{{\rm P}_4}^{{\rm
P}_5} dW^{(1,0,1)}-\int_{{\rm P}_5}^{{\rm P}_6} dW^{(1,1,0)}
-\int_{{\rm P}_6}^{{\rm D}}
dW^{(1,1,1)}\\&=&\frac{4}{3}+\frac{2}{3}\left[\sigma_2(3-\sigma_2^2)+
\sigma_3(3-\sigma_3^2)\right]
\end{eqnarray*}
For instance,
\[
\int_{{\rm D}}^{{\rm P}_1} dW^{(0,0,0)}=\int_0^{\lambda_1^{{\rm
P}_1}}\frac{\lambda_1d\lambda_1}{2\sqrt{1-\lambda_1}}+
\int_{\bar{\sigma}_3^2}^{\lambda_2^{{\rm
P}_1}}\frac{\lambda_2d\lambda_2}{2\sqrt{1-\lambda_2}}+
\int_{\bar{\sigma}_2^2}^{1}\frac{\lambda_3d\lambda_3}{2\sqrt{1-\lambda_3}}
\]
\[
\int_{{\rm P}_1}^{{\rm P}_2} dW^{(0,0,1)}=\int_{\lambda_1^{{\rm
P}_1}}^{\lambda_1^{{\rm
P}_2}}\frac{\lambda_1d\lambda_1}{2\sqrt{1-\lambda_1}}+
\int_{\lambda_2^{{\rm
P}_1}}^{\bar{\sigma}_2^2}\frac{\lambda_2d\lambda_2}{2\sqrt{1-\lambda_2}}-
\int_1^{\bar{\sigma}_2^2}\frac{\lambda_3d\lambda_3}{2\sqrt{1-\lambda_3}},
\]
and so on. It is easy to convince oneself that the sum is
independent of the coordinates $\lambda_i^{{\rm P}_j}$ of the
concrete intersection points P$_j$ and we see that $E({\rm TK3})$
is the same for all the TK3 kinks. The TK3 kink energy is not a
topological quantity because it depends on
$W_1^{(\alpha_1)}(\bar{\sigma}_3^2)$,
$W_2^{(\alpha_2)}(\bar{\sigma}_2^2)$, $W_3^{(\alpha_3)}(1)$, and
not only on the value of $W^{(\alpha_1, \alpha_2, \alpha_3)}$ at
D: $W_1^{(\alpha_1)}(0)$, $W_2^{(\alpha_2)}(\bar{\sigma}_3^2)$,
$W_3^{(\alpha_3)}(\bar{\sigma}_2^2)$.

\subsubsection{Enveloping Kinks.} At the $\lambda_1=0$ face, the
ellipsoid
$\phi_1^2+\frac{\phi_2^2}{{\bar\sigma}_2^2}+\frac{\phi_3^2}{{\bar\sigma}_3^2}=1$
in ${\Bbb R}^3$, we note that the super-potential reduces to:
\begin{equation}
W^{(\alpha_1, \alpha_2 )}(\vec{\mu}) = \frac{1}{3} \left(
(-1)^{\alpha_1} (\mu_1+2) \sqrt{1-\mu_1} +(-1)^{\alpha_2}
(\mu_2+2) \sqrt{1-\mu_2}\right) \label{37}
\end{equation}
Here, we denote by $\lambda_2=\mu_1$, $\lambda_3=\mu_2$ the
components of the two-vector $\vec{\mu}=(\mu_1,\mu_2)$ that
parametrizes the parallelogram ${\bf{\bar P}}_2^\mu$:
${\bar\sigma}_3^2\leq\mu_1\leq{\bar\sigma}_2^2\leq\mu_2\leq 1$.
The reduction of the first-order equations for this
two-dimensional system is:
\begin{eqnarray}
\frac{d\mu_1}{dx}&=& (-1)^{\alpha_1} 2 \frac{
(\mu_1-\bar{\sigma}_2^2)
(\mu_1-\bar{\sigma}_3^2)}{(\mu_1-\mu_2)}\cdot
\sqrt{1-\mu_1}\nonumber\\ \frac{d\mu_2}{dx}&=& (-1)^{\alpha_2} 2
\frac{ (\mu_2-\bar{\sigma}_2^2)
(\mu_2-\bar{\sigma}_3^2)}{(\mu_2-\mu_1)}\cdot
\sqrt{1-\mu_2}\label{38}
\end{eqnarray}
Arguing in a similar vein to that developed in sub-section \S 2.2,
we find that kinks living on the face $\lambda_1=0$ satisfy the
equation:
\begin{equation}
\displaystyle\sum _{a=1}^2
\frac{d\mu_a}{(-1)^{\alpha_a}2(\mu_a-{\bar\sigma}_2^2)
(\mu_a-{\bar\sigma}_3^2)\sqrt{1-\mu_a}}=0 \label{39}
\end{equation}
and have their form factor determined by:
\begin{equation}
\displaystyle\sum _{a=1}^2 \frac{\mu_a
d\mu_a}{(-1)^{\alpha_a}2(\mu_a-{\bar\sigma}_2^2)
(\mu_a-{\bar\sigma}_3^2)\sqrt{1-\mu_a}}=dx . \label{40}
\end{equation}
The integration of equation (\ref{39}) is explicitly shown in the
formula (48) of Reference \cite{1}. These orbits lie on the
$\lambda_1=0$ face of ${\bar {\bf P}}_3(0)$ and are graphically
represented in Figure 4 of \cite{1}. Each member of the family is
a non-topological kink with three non-null components in Cartesian
coordinates. The ${\rm NTK3}$ non-topological kinks are enveloping
kinks in the sense that they live on the face of $\partial{\bar
{\bf P}}_3(0)$; beyond this face in ${\bar {\bf P}}_3(\infty)$,
i.e. for $\lambda_1<0$, there are no kink orbits. The ${\rm NTK3}$
kink form factors are obtained from the integration of (\ref{40}):
\begin{eqnarray}
{\rm
exp}\{2(\gamma_1+x)(\sigma_3^2-\sigma_2^2)\sigma_2\sigma_3\}&=&
\left| \frac{\sqrt{1-\lambda_2}-
\sigma_2}{\sqrt{1-\lambda_2}+\sigma_2}\right|^{\sigma_3{\bar\sigma}_2^2(-1)^\alpha
}\cdot \left| \frac{\sqrt{1-\lambda_2}+\sigma_3}{\sqrt{1-
\lambda_2}-\sigma_3}\right|^{\sigma_2
{\bar\sigma}_3^2(-1)^\alpha}\cdot \nonumber\\ && \left|
\frac{\sqrt{1-\lambda_3}-\sigma_2}{\sqrt{1-\lambda_3}+
\sigma_2}\right|^{\sigma_3{\bar\sigma}_2^2(-1)^\beta}\cdot \left|
\frac{\sqrt{1-\lambda_3}+\sigma_3}{\sqrt{1-\lambda_3}-
\sigma_3}\right|^{\sigma_2 {\bar\sigma}_3^2(-1)^\beta} \label{60}
\end{eqnarray}

A  NTK3 solution starts at $x\to -\infty$ at the vacuum point
D$\equiv (\bar{\sigma}_3^2,\bar{\sigma}_2^2)\in \bar{\bf
P}_2^\mu$, and goes to the $\mu_2=1$ edge, which is crossed, at
$x=x_1$, at point P$_1$. It is thus necessary that both
$\frac{d\mu_1}{dx}$ and $\frac{d\mu_2}{dx}$ should be positive in
the D$\to$P$_1$ step, and hence $\alpha_1$ and $\alpha_2$, by
(\ref{38}), are both 0 in this situation.

From $x_1$ to $x_2$, the kink continues its journey to arrive at
the umbilicus A, with coordinates A$\equiv
(\bar{\sigma}_2^2,\bar{\sigma}_2^2)$. Thus, $\frac{d\mu_1}{dx}$ is
positive and $\frac{d\mu_2}{dx}$ is negative in the P$_1 \to $A
step, and consequently $\alpha_1=0$ and $\alpha_2=1$. The \lq\lq
return" from A to D is made in the opposite sense, with a new
crossing of the BC edge at the point P$_2$. We thus have:
A$\to$P$_2$ ($\alpha_1=1$ and $\alpha_2=0$) and P$_2\to$D
($\alpha_1=1$ and $\alpha_2=1$). We need to solve four different
systems of first-order equations, continuously gluing the
different pieces to describe a complete NTK3 kink. The energy is
thus:
\begin{eqnarray}
& \frac{\lambda^2\sqrt{2}}{m^3}E({\rm NTK3})= -\int_{\rm D}^{{\rm
P}_1} d\left( W_1^{(0)}+W_2^{(0)}\right)- \int_{{\rm P}_1}^{\rm A}
d\left( W_1^{(0)}+W_2^{(1)}\right)\nonumber\\ & -\int_{\rm
A}^{{\rm P}_2} d\left( W_1^{(1)}+W_2^{(0)}\right)- \int_{{\rm
P}_2}^{\rm D} d\left( W_1^{(1)}+W_3^{(1)}\right)= 2 \left(
\sigma_2-\frac{\sigma_2^3}{3}\right) +2 \left(
\sigma_3-\frac{\sigma_3^3}{3}\right)\label{c6finalNTK3}
\end{eqnarray}
Obviously $E({\rm NTK3})$ does not depend on the particular
intersection points P$_1$ and P$_2$, and hence the result is the
same for all the kinks in the family. The NTK3 energy is not a
topological quantity, however, because it depends on the value of
$W$ at the umbilicus point.

\subsubsection{Embedded non-topological Kinks.}

The system can be reduced to the $N=2$ MSTB model twice: in the
$\phi_3=0$-plane and in the $\phi_2=0$-plane. Therefore, all the
NTK2 kinks of the $N=2$ model are embedded in both planes.

The reduction of the system to the $\phi_3=0$-plane in ${\Bbb
R}^3$ occurs in two different faces of ${\bar{\bf P}}_3(0)$:
$\lambda_1=\bar{\sigma}_3^2$ and $\lambda_2=\bar{\sigma}_3^2$. We
shall denote respectively by ${\bf{\bar
P}}_2^{\nu^I}({\bar\sigma}_3^2,{\bar\sigma}_2^2)$ and ${\bf{\bar
P}}_2^{\nu^{II}}(0,{\bar\sigma}_2^2)$ the corresponding
parallelograms:

\noindent ${\bf{\bar
P}}_2^{\nu^I}({\bar\sigma}_3^2,{\bar\sigma}_2^2)$:
${\bar\sigma}_3^2\leq\nu_1^I\leq{\bar\sigma}_2^2\leq\nu_2^I\leq
1$, with $\lambda_1={\bar\sigma}_3^2; \lambda_2=\nu_1^I,
\lambda_3=\nu_2^I, \vec{\nu}^I=(\nu_1^I,\nu_1^I)$

\noindent ${\bf{\bar P}}_2^{\nu^{II}}(0,{\bar\sigma}_2^2)$:
$0\leq\nu_1^{II}\leq{\bar\sigma}_3^2$,
${\bar\sigma}_2^2\leq\nu_2^{II}\leq 1$, with
$\lambda_2={\bar\sigma}_3^2; \lambda_1=\nu_1^{II},
\lambda_3=\nu_2^{II}, \vec{\nu}^{II}=(\nu_1^{II},\nu_1^{II})$

The reduced super-potential is:
\begin{equation}
W^{(\alpha_1, \alpha_2 )}(\vec{\nu}^{I,II}) = \frac{1}{3} \left(
(-1)^{\alpha_1} (\nu_1^{I,II}+2) \sqrt{1-\nu_1^{I,II}}
+(-1)^{\alpha_2} (\nu_2^{I,II}+2) \sqrt{1-\nu_2^{I,II}}\right)
\label{41}
\end{equation}
and the first-order equations in both cases read:
\begin{eqnarray}
\frac{d\nu_1^{I,II}}{dx}&=& (-1)^{\alpha_1} 2 \frac{\nu_1^{I,II}
(\nu_1^{I,II}-\bar{\sigma}_2^2)
}{(\nu_1^{I,II}-\nu_2^{I,II})}\cdot
\sqrt{1-\nu_1^{I,II}}\nonumber\\ \frac{d\nu_2^{I,II}}{dx}&=&
(-1)^{\alpha_2} 2 \frac{
\nu_2^{I,II}(\nu_2^{I,II}-\bar{\sigma}_2^2)}{(\nu_2^{I,II}-\nu_1^{I,II})}\cdot
\sqrt{1-\nu_2^{I,II}}\label{42}
\end{eqnarray}

Analysis of a generic ${\rm NTK2}\sigma_2$ reveals to us the
existence of six different pieces in an ${\rm NTK2}\sigma_2$
orbit. We only write the final result
\begin{eqnarray*}
& \frac{\lambda^2\sqrt{2}}{m^3}\,  E({\rm NTK2}\sigma_2)= -
\int_{{\rm
NTK2}\sigma_2}d(W^{(\alpha_1,\alpha_2,\alpha_3)})=-\int_{\rm
D}^{{\rm P}_1} dW^{(0,0)}(\vec{\nu}^{II}) -\int_{{\rm P}_1}^{{\rm
P}_2} dW^{(0,1)}(\vec{\nu}^{II}) \\ & -\int_{{\rm P}_2}^{{\rm
F}_2} dW^{(0,1)}(\vec{\nu}^{I}) -\int_{{\rm F}_2}^{{\rm P}_3}
dW^{(1,0)}(\vec{\nu}^{I})-\int_{{\rm P}_3}^{{\rm P}_4}
dW^{(1,0)}(\vec{\nu}^{II})-\int_{{\rm P}_4}^{{\rm D}}
dW^{(1,1)}(\vec{\nu}^{II})=
\frac{4}{3}+2\sigma_2(1-\frac{\sigma_2^2}{3})
\end{eqnarray*}
and it is easy to reproduce the study in the $\phi_2=0$-plane,
where we will obtain the energy of the ${\rm NTK2}\sigma_3$
family: $ \frac{\lambda^2\sqrt{2}}{m^3}\,  E({\rm
NTK2}\sigma_3)=\frac{4}{3}+2\sigma_3(1-\frac{\sigma_3^2}{3})$.

\subsubsection{Embedded topological Kinks}

Besides the ${\rm NTK2}\sigma_2$ and ${\rm NTK2}\sigma_3$
families, there are three more embedded kinks from the $N=2$
models, which are topological:
\begin{itemize}
\item TK1. The kink on the $\phi_1$ axis in ${\Bbb R}^3$ is a three-step
orbit running on the edges DF$_3$, F$_3$F$_2$, F$_2$O and back to
D through the same path. Equations (\ref{24}) and their solutions,
centered at the origin, in the three steps are:

\noindent 1. $\lambda_2=\bar{\sigma}_3^2$ and
$\lambda_3=\bar{\sigma}_2^2$. $\frac{d\lambda_1}{dx}=\pm
2\lambda_1 \sqrt{1-\lambda_1},$ $ x\in (-\infty,-\arctanh
\sigma_3]\sqcup [\arctanh \sigma_3,\infty)$. $\lambda_1^{\rm
TK1}(x)=1-\tanh^2x,\quad \lambda_2^{\rm TK1}(x)=\bar{\sigma}_3^2,
\quad \lambda_3^{\rm TK1}(x)=\bar{\sigma}_2^2$

\noindent 2. $\lambda_1=\bar{\sigma}_3^2$ and
$\lambda_3=\bar{\sigma}_2^2$. $\frac{d\lambda_2}{dx}=\pm
2\lambda_2 \sqrt{1-\lambda_2},\quad x\in [-\arctanh
\sigma_3,-\arctanh \sigma_2]\sqcup [\arctanh \sigma_2, \arctanh
\sigma_3]$. $\lambda_1^{\rm TK1}(x)= \bar{\sigma}_3^2,\quad
\lambda_2^{\rm TK1}(x)= 1-\tanh^2x, \quad \lambda_3^{\rm
TK1}(x)=\bar{\sigma}_2^2$

\noindent 3. $\lambda_1=\bar{\sigma}_3^2$ and
$\lambda_2=\bar{\sigma}_2^2$. $\frac{d\lambda_3}{dx}=\pm
2\lambda_3 \sqrt{1-\lambda_3},\quad x\in [-\arctanh
\sigma_2,0]\sqcup [0, \arctanh \sigma_2]$. $\lambda_1^{\rm
TK1}(x)= \bar{\sigma}_3^2,\quad \lambda_2^{\rm TK1}(x)=
\bar{\sigma}_2^2, \quad \lambda_3^{\rm TK1}(x)= 1-\tanh^2x$

The ${\rm TK1}$ kinks are solutions of six different first-order
equations in the intervals: $x\in (\mp\infty,\mp\arctanh
\sigma_3]$, $x\in [\mp\arctanh \sigma_3,\mp\arctanh \sigma_2]$ and
$x\in [\mp\arctanh \sigma_2,0]$.

$E({\rm TK1})=\frac{4m^3}{3\lambda^2\sqrt{2}}$ $\quad$ is not a
topological quantity:
\[
\frac{\lambda^2\sqrt{2}}{m^3}\, E({\rm TK1})= -\int_{\rm D}^{{\rm
F}_3} dW_1^{(0)}-\int_{{\rm F}_3}^{{\rm F}_2}
dW_2^{(0)}-\int_{{\rm F}_2}^{\rm O} dW_3^{(0)}-\int_{\rm O}^{{\rm
F}_2} dW_3^{(1)}-\int_{{\rm F}_2}^{{\rm F}_3}
dW_2^{(1)}-\int_{{\rm F}_3}^{\rm D} dW_1^{(1)}
\]

\item The ${\rm TK2}\sigma_3$ kink. The restriction to the ellipse
$\phi_2=0$, $\phi_1^2+\frac{\phi_3^2}{\bar{\sigma}_3^2}=1$
corresponds in ${\bar{\bf P}}_3(0)$ to two edges of $\partial
{\bar{\bf P}}_3(0)$: DA and AB.

At the DA edge, $\lambda_1=0$ and $\lambda_3=\bar{\sigma}_2^2$,
the system (\ref{38}) reduces to:
\begin{equation}
\frac{d\lambda_2}{dx} = \pm 2
(\lambda_2-\bar{\sigma}_3^2)\sqrt{1-\lambda_2}
\end{equation}
with the solution, centered at the origin $x_0=0$:
\begin{equation}
\lambda_1^{{\rm TK2}\sigma_3}(x)=0,\quad \lambda_2^{{\rm
TK2}\sigma_3}(x)=1-\sigma_3^2 \tanh^2 (\sigma_3 x),\quad
\lambda_3^{{\rm TK2}\sigma_3}(x)=\bar{\sigma}_2^2
\end{equation}
for $x\in (-\infty, \frac{-1}{\sigma_3}
\arctanh\frac{\sigma_2}{\sigma_3} ]\sqcup [\frac{1}{\sigma_3}
\arctanh \frac{\sigma_2}{\sigma_3},\infty)$. In the second step,
the AB edge, equations (\ref{38}) again reduce to a single
differential equation: if $\lambda_1=0$ and
$\lambda_2=\bar{\sigma}_2^2$, then
\begin{equation}
\frac{d\lambda_3}{dx}=\pm
2(\lambda_3-\bar{\sigma}_3^2)\sqrt{1-\lambda_3}
\end{equation}
has the solution
\begin{equation}
\lambda_1^{{\rm TK2}\sigma_3}(x)=0,\quad \lambda_2^{{\rm
TK2}\sigma_3}(x)= \bar{\sigma}_2^2,\quad \lambda_3^{{\rm
TK2}\sigma_3}(x)=1-\sigma_3^2 \tanh^2 (\sigma_3 x)
\end{equation}
for $x\in \left[ \frac{-1}{\sigma_3}
\arctanh\frac{\sigma_2}{\sigma_3} , \frac{1}{\sigma_3} \arctanh
\frac{\sigma_2}{\sigma_3}\right]$. The energy is thus:
\begin{eqnarray*}
&\frac{\lambda^2\sqrt{2}}{m^3} E({\rm TK2}\sigma_3)= \int_{{\rm
TK2}\sigma_3}d(W_2^{(\alpha_2)}+W_3^{(\alpha_3)})=\\ &=-\int_{\rm
D}^{\rm A} dW_2^{(0)}+\int_{\rm A}^{\rm B} dW_3^{(0)}+ \int_{\rm
B}^{\rm A} dW_3^{(1)}+\int_{\rm A}^{\rm D} dW_2^{(1)}=
2\sigma_3\left(1-\frac{\sigma_3^2}{3})\right)
\end{eqnarray*}
The two steps are continuously sewn at the umbilicus point
A$\equiv(0,\bar{\sigma}_2^2,\bar{\sigma}_2^2)$. Note that the
${\rm TK2}\sigma_3$ kinks are solutions of four $\underline{{\rm
different}}$ first-order equations on the intervals
$x\in(\mp\infty,\frac{\mp 1}{\sigma_3}{\rm
arctanh}\frac{\sigma_2}{\sigma_3}]$ and $x\in[\frac{\mp
1}{\sigma_3}{\rm arctanh}\frac{\sigma_2}{\sigma_3},\frac{\pm
1}{\sigma_3}{\rm arctanh}\frac{\sigma_2}{\sigma_3}]$ glued at the
umbilicus point A. $E({\rm TK2}\sigma_3)$ is not a topological
quantity.

\item The ${\rm TK2}\sigma_2$ kink. The restriction to the ${\rm DC}$ edge,
$\lambda_1=0$ and $\lambda_2={\bar\sigma}_3^2$, corresponds, in
${\Bbb R}^3$, to the ellipse $\phi_3=0$ and
$\phi_1^2+\frac{\phi_2^2}{\bar{\sigma}_2^2}=1$. System (\ref{38})
reduces to:
\begin{equation}
\frac{d\lambda_3}{dx}=(-1)^\beta(\lambda_3-{\bar\sigma}_2^2)\sqrt{1-\lambda_3}.
\end{equation}
The kink form factor is thus,
\[
\lambda_1^{{\rm TK2}\sigma_2}(x)=0 , \hspace{0.5cm}
\lambda_2^{{\rm TK2}\sigma_2}(x)={\bar\sigma}_3^2 , \hspace{0.5cm}
\lambda_3^{{\rm TK2}\sigma_2}(x)=1-{\bar\sigma}_2^2{\rm
tanh}^2(\sigma_2x)
\]
and the energy reads:
\[
\frac{\lambda^2\sqrt{2}}{m^3} E({\rm TK2}\sigma_2)=-\int_{{\rm
TK2}\sigma_2}dW_3^{(\beta)}=-\left[ \int_{\rm D}^{\rm C}
dW_3^{(0)}+\int_{\rm C}^{\rm D} dW_3^{(1)}\right]=
2\sigma_2\left(1-\frac{\sigma_2^2}{3}\right)
\]

Note that $W_3^{(\beta)}(C)=0$, $\forall \beta=0,1$, and hence the
energy of the TK2$\sigma_2$ kink depends only on the value of
$W_3^{(0)}$ and $W_3^{(1)}$ at the vacuum D. $E$(TK2$\sigma_2$) is
thus a topological quantity and in consequence TK2$\sigma_2$ is
the absolute minimum of the functional $E$; necessarily, the
TK2$\sigma_2$ kink is a stable solution.

\end{itemize}

The rest of the kinks analyzed are not absolute minima of $E$,
although they could be local minima. We shall see in section \S 4
that this is not the case and that all of them are unstable
critical curves of $E$. According to the usual classification
between BPS and non-BPS solitary waves in supersymmetric theories,
we have shown that the only BPS solution of the equations of the
system is the TK2$\sigma_2$ kink.

\section{The moduli space of kinks}

\subsection{kink-energy sum rules}

The energies of the different kinks that we have calculated in the
previous section are related in several ways. By inspection, we
see the energy of a generic TK3 kink as the sum of the energies of
embedded and enveloping kinks. There are four primary sum rules:

\noindent 1)
\begin{equation}
\fbox{$E({\rm TK3})= E({\rm NTK2}\sigma_3)+ E({\rm
TK2}\sigma_2)$}\label{65}
\end{equation}

\noindent 2)
\begin{equation}
\fbox{$E({\rm TK3})= E({\rm NTK2}\sigma_2) + E({\rm
TK2}\sigma_3)$}\label{66}
\end{equation}

\noindent 3)
\begin{equation}
\fbox{$E({\rm TK3})= E({\rm NTK3})+ E({\rm TK1})$}\label{67}
\end{equation}

\noindent 4)
\begin{equation}
\fbox{$E({\rm NTK3})= E({\rm TK2}\sigma_3)+ E({\rm
TK2}\sigma_2)$}\label{68}
\end{equation}
As a consequence, we also find:
\begin{equation}
E({\rm TK3})= E({\rm TK1})+ E({\rm TK2}\sigma_2)+ E({\rm
TK2}\sigma_3)\label{68b}
\end{equation}
The old kink-energy sum rules, already appearing in the $N=2$
model \cite{8} \cite{9} and involving only embedded kinks, are
also satisfied:
\begin{equation}
E({\rm NTK2}\sigma_3)= E({\rm TK2}\sigma_3)+ E({\rm TK1});\quad
E({\rm NTK2}\sigma_2)= E({\rm TK2}\sigma_2)+ E({\rm
TK1})\label{68a}.
\end{equation}

\subsection {The variety of kink trajectories: Singular limits}
In this sub-section we shall show how the kink solutions of the
dimensionally reduced equations are singular limits of the generic
${\rm TK3}$ kinks obtained by letting $\gamma_2$ and $\gamma_3$ go
to $\pm\infty$ in formulae (35) and (36) of Reference \cite{1}.
This circumstance fully explains the origin of the kink-energy sum
rules: suitable combinations of enveloping and embedded kinks
arise at the boundary of the families of generic kinks.

In order to describe the process of reaching the singular limits
in a complete way , it is convenient to look at the manifold of
TK3 kinks as the set of solutions of a re-shuffling of the
equations (\ref{26}) and (\ref{27}). Two equivalent equations are
obtained through multiplication of (\ref{27}); first by
${\bar\sigma}_3^2$, then by ${\bar\sigma}_2^2$, and subtraction of
(\ref{26}) from both expressions. The solutions to this equivalent
system are:
\begin{eqnarray}
e^{2\sigma_2 \bar{\sigma}_2^2\gamma} &=& \left( \left|
\frac{\sqrt{1-\lambda_1}-1}{\sqrt{1-\lambda_1}+1}\right|^{\sigma_2}
\cdot \left|
\frac{\sqrt{1-\lambda_1}+\sigma_2}{\sqrt{1-\lambda_1}-\sigma_2}\right|\right)^{(-1)^{\alpha_1}}\cdot\nonumber
\\ && \left( \left|
\frac{\sqrt{1-\lambda_2}-1}{\sqrt{1-\lambda_2}+1}\right|^{\sigma_2}
\cdot \left|
\frac{\sqrt{1-\lambda_2}+\sigma_2}{\sqrt{1-\lambda_2}-\sigma_2}\right|\right)^{(-1)^{\alpha_2}}\cdot\label{69}\\
&& \left( \left|
\frac{\sqrt{1-\lambda_3}-1}{\sqrt{1-\lambda_3}+1}\right|^{\sigma_2}
\cdot \left|
\frac{\sqrt{1-\lambda_3}+\sigma_2}{\sqrt{1-\lambda_3}-\sigma_2}\right|\right)^{(-1)^{\alpha_3}}\nonumber
\end{eqnarray}
and
\begin{eqnarray}
e^{2\sigma_3 \bar{\sigma}_3^2\bar{\gamma}}&=& \left( \left|
\frac{\sqrt{1-\lambda_1}-1}{\sqrt{1-\lambda_1}+1}\right|^{\sigma_3}
\cdot \left|
\frac{\sqrt{1-\lambda_1}+\sigma_3}{\sqrt{1-\lambda_1}-\sigma_3}\right|\right)^{(-1)^{\alpha_1}}\cdot\nonumber
\\ && \left( \left|
\frac{\sqrt{1-\lambda_2}-1}{\sqrt{1-\lambda_2}+1}\right|^{\sigma_3}
\cdot \left|
\frac{\sqrt{1-\lambda_2}+\sigma_3}{\sqrt{1-\lambda_2}-\sigma_3}\right|\right)^{(-1)^{\alpha_2}}\cdot\label{70}\\
&& \left( \left|
\frac{\sqrt{1-\lambda_3}-1}{\sqrt{1-\lambda_3}+1}\right|^{\sigma_3}
\cdot \left|
\frac{\sqrt{1-\lambda_3}+\sigma_3}{\sqrt{1-\lambda_3}-\sigma_3}\right|\right)^{(-1)^{\alpha_3}}\nonumber
\end{eqnarray}
where the \lq\lq new" integration constants are: $\gamma=\gamma_3
\bar{\sigma}_3^2 -\gamma_2$ and $\bar{\gamma}=\gamma_3
\bar{\sigma}_2^2 -\gamma_2$.

All the generic kink orbits are described analytically  by
(\ref{69}) and (\ref{70}). The advantage of using $\gamma$ and
$\bar{\gamma}$ in the parametrization of the TK3 family is the
following: in terms of the constants $\gamma$ and $\bar{\gamma}$
one determines the intersection points of a generic TK3 orbit with
the edges of ${\bar {\bf P}}_3(0)$
$\lambda_1=\lambda_2={\bar\sigma}_3^2$, the ellipse (\ref{33}) in
Cartesian coordinates, and $\lambda_2=\lambda_3={\bar\sigma}_2^2$,
the hyperbola (\ref{34}), which are precisely the lines along
which the super-potential is non-differentiable. If
$\tilde{\lambda}_3$ is a root of the equation:
\begin{equation}
e^{2\sigma_2 \bar{\sigma}_2^2 \gamma} \, =\, \left( \left|
\frac{\sqrt{1-\lambda_3}-1}{\sqrt{1-\lambda_3}+1}\right|^{\sigma_2}
\cdot \left|
\frac{\sqrt{1-\lambda_3}+\sigma_2}{\sqrt{1-\lambda_3}-\sigma_2}\right|
\right)^{(-1)^{\alpha_3}}\label{Ia}
\end{equation}
the point $(\lambda_1,\lambda_2,\lambda_3) = (\bar{\sigma}_3^2,
\bar{\sigma}_3^2,\tilde{\lambda}_3)$ is the intersection point of
the TK3 orbit with the edge F$_1$F$_3$ in ${\bar {\bf P}}_3(0)$
(the ellipse (\ref{33}) in ${\Bbb R}^3$). There are two
$\tilde{\lambda}_1^{\pm}$ roots of
\begin{equation}
e^{2\sigma_3 \bar{\sigma}_3^2 \bar{\gamma}} \, =\, \left( \left|
\frac{\sqrt{1-\lambda_1}-1}{\sqrt{1-\lambda_1}+1}\right|^{\sigma_3}
\cdot \left|
\frac{\sqrt{1-\lambda_1}+\sigma_3}{\sqrt{1-\lambda_1}-\sigma_3}\right|
\right)^{(-1)^{\alpha_1}}\quad ,\label{Iia}
\end{equation}
depending on the choices of $\alpha_1$, which determine the two
intersection points of the TK3 kink with the edge AF$_2$ in
${\bar{\bf P}}_3(0)$ (the hyperbola (\ref{34}) in ${\Bbb R}^3$).
Thus, the analytical parametrization by $(\gamma,\bar{\gamma})$ of
the TK3 family acquires a geometric meaning. A choice of
$\bar{\gamma}$ fixes the intersection point of a given ${\rm TK3}$
with the edge AF$_2$. There are, however, infinitely many ${\rm
TK3}$ orbits meeting at this point in AF$_2$; this congruence
comes from each point in F$_1$F$_3$ and is parametrized by
$\gamma\in {\Bbb R}$. Conversely, when $\bar{\gamma}$ varies in
${\Bbb R}$ a family of TK3 orbits is described that crosses every
point in AF$_2$ and meets at a single point at the edge
F$_1$F$_3$, characterized by a fixed value of $\gamma$. The
corresponding conics in ${\Bbb R}^3$, the ellipse (\ref{33}), and
the hyperbola (\ref{34}) are lines where the gradient flow of the
super-potential is undefined: a one-parameter family of TK3 orbits
touching all points of the ellipse (\ref{33}) meets at a single
point of the hyperbola (\ref{34}) and vice-versa.

It should be stressed that the boundary points A, F$_1$, F$_2$ and
F$_3$ are excluded because they are reached when either $\gamma$
or $\bar{\gamma}$ goes to $\pm \infty$, and for these values there
are no generic kink orbits. The question arises: are there any
kink orbits when $\gamma$ or $\bar{\gamma}$  are $\pm \infty$?, or
alternatively, when $\gamma_2$ or $\gamma_3$ are $\pm \infty$?  We
shall show that this is indeed the case, at the same time
justifying the kink-energy sum rules.

In the process of taking $\gamma\to \pm \infty$ or
$\bar{\gamma}\to \infty$ (or the corresponding $\gamma_2$ and/or
$\gamma_3\to \pm\infty$) different singular solutions appear,
depending on the choice of $\alpha_1, \alpha_2$ and $\alpha_3$ in
equations (\ref{69}) and (\ref{70}). For instance, taking
$\gamma\to -\infty$ in (\ref{69}) leads us to $\lambda_1\to 0$ if
we consider $\alpha_1=0$, but it is also compatible with
$\lambda_2\to \bar{\gamma}_2^2$ if $\alpha_2=1$ with no
restrictions to the value of $\alpha_1$, and so on. In Reference
\cite{7} the singular limits have been carefully explored, looking
at all the combinations of the different signs allowed. We here
summarize only the final results, because the very technical
details add nothing further to our conceptual knowledge of the
subject.

The five possible singular limits in the TK3 kink space are:

\noindent $\bullet$ $\gamma\to \pm \infty$, $\bar{\gamma}$ finite.
As one can check in Appendix A, taking $\gamma\to \pm \infty$,
$\bar{\gamma}$ remaining finite, leads us, in a non-trivial way,
to the combination of an ${\rm NTK2}\sigma_3$ (determined by the
particular value of $\bar{\gamma}$) plus the ${\rm TK2}\sigma_2$.
Later taking the $\bar{\gamma}\to \pm \infty$ limit, one arrives
at the combination of orbits ${\rm TK2}\sigma_3$+ ${\rm
TK2}\sigma_2$+ TK1.

\noindent $\bullet$ $\bar{\gamma}\to \pm \infty$, finite $\gamma$.
In a similar way, it is possible to show that this limit is
reached at the combination of one ${\rm NTK2}\sigma_2$
(parametrized by $\gamma$) plus the ${\rm TK2}\sigma_3$ orbit.
Later taking the $\gamma\to \pm \infty$ limit leads to ${\rm
TK2}\sigma_3$+ ${\rm TK2}\sigma_2$+ TK1.

\noindent $\bullet$ $\gamma_3\to \pm \infty$, finite $\gamma_2$.
In this case, the result is an NTK3 orbit (determined by the value
of $\gamma_2$) plus the TK1 orbit. Later passing to the
$\gamma_2\to \pm \infty$ limit produces the combination ${\rm
TK2}\sigma_3$+ ${\rm TK2}\sigma_2$+ TK1.

\noindent $\bullet$ $\gamma_2\to \pm \infty$, finite $\gamma_3$.
This situation is in some sense the most singular; regardless of
the particular value of the constant $\gamma_3$, the limit is
always the combination ${\rm TK2}\sigma_3$+ ${\rm TK2}\sigma_2$+
TK1.

\noindent $\bullet$ Any other possibility of going to infinity in
the space of parameters of the TK3 family, without the
restrictions detailed in the previous cases, leads to the \lq\lq
most singular" combination: ${\rm TK2}\sigma_3$+ ${\rm
TK2}\sigma_2$+ TK1.

\subsection{Moduli space structure}

The moduli space of kinks is defined as the space of solitary wave
solutions of the field equations modulo the action of the symmetry
group $G={\Bbb Z}^{\times 3}$ of the system. The action of the
group $G$ on the different kinds of kinks advanced in the
Introduction is now clear after the analysis of the solutions
discussed in the previous Sections. The sub-space of \lq\lq
points" that are not fixed under the action of any non-trivial
sub-group of $G$ is formed precisely by the solutions of
(\ref{69})-(\ref{70}) for finite values of $\gamma$ and
$\bar{\gamma}$: at any point $(\gamma,\bar{\gamma})\in {\Bbb
R}^2-\partial {\Bbb R}^2$ there is associated a generic TK3 kink
trajectory in elliptic coordinates that corresponds to eight TK3
kinks in Cartesian space. Therefore, the moduli space of TK3 kinks
in the deformed linear $O(3)$-sigma model is the open plane
parametrized by the $(\gamma,\bar{\gamma})$ integration constants.
The aim of this Section is to describe how the other kinks of the
system enter this moduli space. Any \lq\lq member" of the moduli
space must have the same energy as any other. Thus, we expect that
the other kinks enter the moduli space of TK3 kinks assembled in
such a way that the kink-energy sum rules will be saturated.

The embedded topological and non-topological kinks and the NTK3
enveloping kinks live at the boundary of the TK3 moduli space
${\Bbb R}^2$ and provide a compactification of this space. The
kind of compactification depends on how one goes to infinity in
${\Bbb R}^2$ and how one chooses a particular degeneration of one
${\rm TK}3$ kink to a specific assembly of the other kinks. To
simplify the cumbersome labeling, we shall denote the different
kinds of kinks according to the following Table:

\begin{center}
\begin{tabular}{|l|l|}
\hline {\small TK3} &  ${\rm T}_3$
\\ \hline {\small NTK3+TK1} &  ${\rm N}_3{\rm T}_1$
\\ \hline {\small NTK2$\sigma_3$+TK2$\sigma_2$} &  ${\rm N}_2^{\sigma_3}{\rm T}_2^{\sigma_2}$
\\ \hline {\small NTK2$\sigma_2$+TK2$\sigma_3$} &  ${\rm N}_2^{\sigma_2}{\rm T}_2^{\sigma_3}$
\\ \hline {\small TK2$\sigma_2$+TK2$\sigma_3$+TK1} &  ${\rm T}_2^{\sigma_2}{\rm T}_2^{\sigma_3}
{\rm T}_1$\\ \hline
\end{tabular}
\end{center}

Only three \lq\lq paths" to infinity will be important in
developing degenerate Morse theory for the configuration space:
\begin{description}
    \item{-a)}  Taking
    the limits $\Lambda \rightarrow \pm \infty$ in the family of straight lines
    $\gamma=\Lambda$, $\Lambda\in {\Bbb R}$, the moduli space of
    N$_2^{\sigma_3}$T$_2^{\sigma_2}$ singular kink
     configurations is found. The plane becomes an infinite cylinder
     through identification of the $\Lambda=\pm\infty$ lines.
     Moreover, all the points in the $\bar{\gamma}=\pm\infty$
     circles give the same T$_2^{\sigma_2}$T$_2^{\sigma_3}$T$_1$
     configuration -skipping the path which reaches these circles as described
     in b)- and must be identified to a point. The addition of
     two  N$_2^{\sigma_3}$T$_2^{\sigma_2}$ families amounts to
     compactifying the TK3 moduli space to a $S^2$ sphere:
     $\bar{{\cal M}}_{{\rm T}_3}^{(1)}={\cal M}_{{\rm T}_3}\sqcup
      2{\cal M}_{{\rm N}_2^{\sigma_3}{\rm T}_2^{\sigma_2}}\cong
     {S}^2$.

    \item{-b)} The above behaves in
    exactly the same way when the $\gamma=\Lambda$ are replaced by the $\bar{\gamma}=\Lambda$
    lines. The N$_2^{\sigma_2}$T$_2^{\sigma_3}$ moduli space
    arises on the $\Lambda=\pm\infty$ lines. The gluing of these
    two lines provides an infinite cylinder, and the
    identification of the two $\gamma=\pm\infty$ circles at
    infinity to a point, the T$_2^{\sigma_2}$T$_2^{\sigma_3}$T$_1$
    configuration, leads to the two-sphere:
    $\bar{{\cal M}}_{{\rm T}_3}^{(2)}={\cal M}_{{\rm T}_3}\sqcup
      2{\cal M}_{{\rm N}_2^{\sigma_2}{\rm T}_2^{\sigma_3}}\cong
     {S}^2$. Here, we skip the path to the $\gamma=\pm\infty$ circles
     described in a).
    \item{-c)}  The third interesting possibility is to take the
    $\Lambda\rightarrow\pm\infty$ limit in the family of straight
    lines: $\bar{\gamma}=\gamma -\Lambda$. Note that
    $\gamma_3=$ constant on these lines. We find the N$_3$T$_1$ moduli space on
    both limits and the identification of these two lines leads to a third infinite cylinder.
     Again, the $\gamma_3=\pm\infty$ circles must be identified
     because all their points are the
     T$_2^{\sigma_2}$T$_2^{\sigma_3}$T$_1$ configuration, and hence:$\bar{{\cal M}}_{{\rm T}_3}^{(3)}={\cal M}_{{\rm T}_3}\sqcup
      2{\cal M}_{{\rm N}_3{\rm T}_1}\cong
     {S}^2$.
\end{description}

\section{Kink Stability}

Let $\vec{\psi}(x)$ be a path in a Riemannian manifold
$({M}^3,g)$. Using the language of variational calculus, let us
denote by $\vec{\psi}(x,\xi)$, $\xi\in
(-\varepsilon,\varepsilon)$, a proper variation of $\vec{\psi}$
and let $\dot{\vec{\psi}}(x,\xi)$, $\vec{V}(x,\xi)$ be the vector
fields
\[
\dot{\vec{\psi}}(x,\xi)=\frac{\partial\vec{\psi}}{\partial
x}(x,\xi)\hspace{0.5cm} ,\hspace{0.5cm}
\vec{V}(x,\xi)=\frac{\partial\vec{\psi}}{\partial
\xi}(x,\xi)\qquad .
\]
The Hessian quadratic form, the second variation of the energy
functional,
\begin{equation}
E[\vec{\psi}]=\int dx \left(
\frac{1}{2}\left\langle\frac{d\vec{\psi}}{dx},\frac{d\vec{\psi}}{dx}\right\rangle
+ U(\vec{\psi})\right)\qquad ,\label{eq:enerr}
\end{equation}
at a \lq\lq critical point" - a kink solution $\vec{\psi}_K$ for
which the first variation of $E[\vec{\psi}]$ is zero-, is
\begin{equation}
{\cal H}=\int_{\vec{\psi}_{\rm K}} dx \left<\Delta \vec{V},\vec{V}
\right> =\int_{\vec{\psi}_{\rm K}} dx \left<-\frac{D^2 \vec{V}
}{dx^2}-K_{\dot{\vec{\psi}}_{\rm K}}(\vec{V})+\nabla_{\vec{V}}
\vec{\rm grad}(U), \vec{V}\right>
\end{equation}
Here, $\frac{D\vec{V}}{dx}$ is the covariant derivative along the
kink orbit, $K_{\dot{\vec{\psi}}_{\rm
K}}(\vec{V})=R(\dot{\vec{\psi}}_{\rm K} ,\vec{V}
)\dot{\vec{\psi}}_{\rm K}$ is the sectional curvature defined in
terms of the curvature tensor $R$ and $\nabla_{\vec{V}}\vec {\rm
grad}(U)$ is the covariant derivative of $\vec{\rm grad}\,U$ in
the direction of $\vec{V}$. In a local coordinate system in
${M}^3$, the differential operator $\Delta$ reads
\[
\Delta_b^a = - \frac{D^2}{d x^2} \,
\delta_b^a-[K_{\dot{\vec{\psi}}_{\rm K}}]_b^a+ {\cal
U}_b^a(\vec{\psi}_{\rm K})
\]
\[
\frac{DV^a}{dx}=\frac{dV^a}{dx}+\Gamma^a_{bc}(\vec{\psi}_K)\frac{
d\psi^b_K}{dx}V^c\, , [K_{\dot{\psi}_{\rm
K}}]_b^a=R^a_{cbd}\dot{\psi}_{\rm K}^c\dot{\psi}_{\rm K}^d , \quad
{\cal U}_b^a(\vec{\psi}_{\rm K})=\frac{\partial^2 U}{\partial
\psi_a\partial\psi^b}(\vec{\psi}_{\rm
K})-\Gamma^a_{cb}\frac{\partial U}{\partial\psi_c}(\vec{\psi}_{\rm
K})
\]
We shall consider $\vec{\psi}$ as the static scalar field either
in Cartesian -$\vec{\phi}(x)$- or in elliptic -$\vec{\lambda}(x)$-
coordinates. Therefore, ${M}^3$ is the Euclidean ${\Bbb R}^3$
space in the first case and ${\bf P}_3(\infty)$ with the metric
induced by the change of coordinates, see Section \S 2.2 , in the
later case. Either way, the stability of kink solutions against
small deformations is encoded in the spectrum of $\Delta$.

\subsection{Jacobi fields and the Jacobi Theory}
Even without knowing the analytical expression of $\Delta$, we can
state a general fact about its spectrum: if there exists a kink
family $\vec{\psi}_{\rm K}=\vec{\psi}_{\rm K}(x,\gamma)$ of
critical points of $E[\vec{\psi}]$ characterized by the value of
the parameter $\gamma$, then the vector field $\frac{\partial
\vec{\psi}_{\rm K}}{\partial \gamma}$ is a Jacobi field, i.e. it
belongs to the kernel of $\Delta$,
\[
\left\langle \Delta \frac{\partial \vec{\psi}_{\rm K}}{\partial
\gamma},\frac{\partial \vec{\psi}_{\rm K}}{\partial
\gamma}\right\rangle=\frac{\partial }{\partial \gamma}\left\langle
-\frac{D}{\partial x} \dot{\vec{\psi}}_{\rm K}+\vec{\rm grad}
U(\vec{\psi}_{\rm K}),\frac{\partial \vec{\psi}_{\rm K}}{\partial
\gamma}\right\rangle - \left\langle -\frac{D}{\partial x}
\dot{\vec{\psi}}_{\rm K}+\vec{\rm grad} U(\vec{\psi}_{\rm
K}),\frac{D}{\partial \gamma} \frac{\partial \vec{\psi}_{\rm
K}}{\partial \gamma }\right\rangle=0
\]

Any generic ${\rm TK}3$ kink therefore gives rise to three Jacobi
fields: $\frac{\partial\vec{\psi}_{{\rm TK}3}}{\partial\gamma_1}$
is tangent to the orbit and is usually termed as the translational
mode. $\frac{\partial\vec{\psi}_{{\rm TK}3}}{\partial\gamma_2}$
and $\frac{\partial\vec{\psi}_{{\rm TK}3}}{\partial\gamma_3}$,
however, have non-zero components in the orthogonal sub-space to
the trajectory in ${\cal C}$ and obey zero \lq\lq particle energy"
fluctuations along the kink moduli space. Both the embedded and
enveloping non-topological kinks form two-parametric families of
solutions. Besides the translational mode, all of them present a
second Jacobi field coming from the partial derivative of the
field configuration with respect to the second integration
constant.

According to the theory elaborated by Jacobi in the 19th century,
see \cite{16}, Jacobi fields contain a lot of information about
the stability of the trajectories to which they are associated.
The application to our problem is as follows:

\lq\lq A kink trajectory is unstable if there is a non-trivial
Jacobi field that vanishes at the endpoints of the real half-line:
 $(-\infty,x_0]$". Essentially the argument is based on the
following idea: the projection of $\Delta$ in the direction of the
Jacobi field is a Schrodinger operator. The Jacobi field itself
belongs to the kernel of this operator but has at least one node:
it cannot be the ground state.
 Therefore, the strategy to determine whether or not a kink
orbit is stable is based on the study of the features of the
Jacobi fields along these trajectories.
\begin{enumerate}
\item {\bf Non-topological kinks}

We  first focus on the {\bf NTK} kinks. The three types can be
described in a unified way by using the two-dimensional elliptic
variables: $c\leq\mu_1\leq\bar{\sigma^2}\leq\mu_2\leq 1$.
Appropriate re-scalings and choices of the constants $c$ and
$\bar{\sigma}^2$ lead to the elliptic coordinates which work for
the ${\rm NTK}3$, ${\rm NTK}2\sigma^2_3$, and ${\rm
NTK}2\sigma^2_2$ kinks. The kink orbit and the kink form factor
are respectively determined by the equations:
\begin{equation}
\left(\left|
\frac{\sqrt{1-\mu_1}-\sigma}{\sqrt{1-\mu_1}+\sigma}\right| \left|
\frac{\sqrt{1-\mu_1}+1}{\sqrt{1-\mu_1}-1}\right|^{\sigma}\right)^{(-1)^{\alpha_1}}
 \left( \left|
\frac{\sqrt{1-\mu_2}-\sigma}{\sqrt{1-\mu_2}+\sigma}\right| \left|
\frac{\sqrt{1-\mu_2}+1}{\sqrt{1-\mu_2}-1}\right|^{\sigma}\right)^{(-1)^{\alpha_2}}=e^{2\sigma\bar{\sigma}^2
\gamma_2}\label{orbitasNTK2}
\end{equation}
\begin{equation}
\left| \frac{
\sqrt{1-\mu_1}-\sigma}{\sqrt{1-\mu_1}+\sigma}\right|^{\bar{\sigma}^2{(-1)^{\alpha_1}}}
\cdot \left|
\frac{\sqrt{1-\mu_2}-\sigma}{\sqrt{1-\mu_2}+\sigma}\right|^{\bar{\sigma}^2(-1)^{\alpha_2}}
=e^{2\sigma(x+\gamma_1)}\label{factorformaNTK2}
\end{equation}
It is not possible to invert the system
(\ref{orbitasNTK2})-(\ref{factorformaNTK2}) and write $\mu_1$ and
$\mu_2$ explicitly as elementary transcendental functions of $x$ ;
henceforth, one cannot explicitly write the Hessian operator for
these configurations. To identify the orthogonal Jacobi field,
however, we perform the implicit derivation of (\ref{orbitasNTK2})
and (\ref{factorformaNTK2}) with respect to $\gamma_2$ and solve
the linear system to find the components
$\frac{\partial\mu_1^{{\rm NTK}}}{\partial\gamma_2}$,
$\frac{\partial\mu_2^{{\rm NTK}}}{\partial\gamma_2}$ of the Jacobi
field.

Before of doing so, we must first address a very subtle point: all
the kink orbits -the solutions of (\ref{orbitasNTK2})
$\forall\gamma_2$- pass through the vacuum point
$(\bar{\mu}_1=0,\bar{\mu}_2=\bar{\sigma}^2)$ in the
$\alpha_1=\alpha_2$ steps. There is only one other point crossed
by all the kink orbits: the focus
$(\bar{\mu}_1=\bar{\sigma}^2,\bar{\mu}_2=\bar{\sigma}^2)$ is
reached by all the kink orbits in the $\alpha_1=0,\alpha_2=1$ step
and left in the $\alpha_1=1,\alpha_2=0$ step. Note that the focus
becomes the umbilicus point for the ${\rm NTK}3$ kinks and the
appropriate focus of the ellipsoid for the ${\rm NTK}2\sigma_2$ or
${\rm NTK}2\sigma_3$ kinks. Given a kink orbit
$(\bar{\mu}_1(x,\gamma_2),\bar{\mu}_2(x,\gamma_2))$, near the
focus (\ref{orbitasNTK2}) becomes :
\begin{equation}
\lim_{(\bar{\mu}_1,\bar{\mu}_2)\to
(\bar{\sigma}^2,\bar{\sigma}^2)}\frac{\sqrt{1-\bar{\mu}_1}-\sigma}{\sqrt{1-\bar{\mu}_2}-\sigma}
=e^{\pm 2\sigma\bar{\sigma}^2 \gamma_2} \qquad \label{limit}
\end{equation}
the +/- signs standing for the steps arriving at/departing from
the focus. Taking the same limit in equation
(\ref{factorformaNTK2}) and using (\ref{limit}), one immediately
sees that the focus is reached at the \lq\lq instant"
$x_0=-\gamma_1+\bar{\sigma}^4\gamma_2$. Here, the subtlety is:
different kink trajectories reach the focus at different times.

Defining $\bar{\gamma}_1=\gamma_1-\bar{\sigma}^4 \gamma_2$
equation (\ref{factorformaNTK2}) is replaced by the equivalent
equation :
\begin{equation}
\left|
\frac{\sqrt{1-\mu_1}-1}{\sqrt{1-\mu_1}+1}\right|^{\bar{\sigma}^2(-1)^{\alpha}_1}
\cdot \left|
\frac{\sqrt{1-\mu_2}-1}{\sqrt{1-\mu_2}+1}\right|^{\bar{\sigma}^2(-1)^{\alpha_2}}
=e^{\pm 2\sigma(x+\bar{\gamma}_1)}\label{factorformaNTK2bis}
\end{equation}
obtained by multiplying, member by member, equation
(\ref{factorformaNTK2}) by the $\bar{\sigma}^2$ power of equation
(\ref{orbitasNTK2}). By identical arguments as above it can be
shown that the trajectories that solve (\ref{orbitasNTK2}) and
(\ref{factorformaNTK2bis}) reach the focus at the same \lq\lq
instant": $x_0=-\bar{\gamma}_1$. Thus, we have shown that the
focus is a conjugate point to the vacuum.

We now use the derivatives of (\ref{orbitasNTK2}) and
(\ref{factorformaNTK2bis}) with respect to $\gamma_2$ to find the
Jacobi field orthogonal to each kink trajectory
$(\bar{\mu}_1(x,\bar{\gamma}_1,\gamma_2),\bar{\mu}_2(x,\bar{\gamma}_1,\gamma_2)$:
if $\alpha_1\neq\alpha_2$,
\begin{equation}
\frac{-(-1)^{\alpha_1}}{\bar{\mu}_1\sqrt{1-\bar{\mu}_1}(\bar{\sigma}^2-\bar{\mu}_1)}\frac{\partial
\bar{\mu}_1}{\partial
\gamma_2}-\frac{(-1)^{\alpha_2}}{\bar{\mu}_2\sqrt{1-\bar{\mu}_2}(\bar{\sigma}^2-\bar{\mu}_2)}\frac{\partial
\bar{\mu}_2}{\partial \gamma_2}=2 \label{eq1}
\end{equation}
\begin{equation}
\frac{(-1)^{\alpha_1}}{\bar{\mu}_1\sqrt{1-\bar{\mu}_1}}\frac{\partial
\bar{\mu}_1}{\partial
\gamma_2}+\frac{(-1)^{\alpha_2}}{\bar{\mu}_2\sqrt{1-\bar{\mu}_2}}\frac{\partial
\bar{\mu}_2}{\partial \gamma_2}=0\label{eq2}
\end{equation}

The solution of the linear system of algebraic equations
(\ref{eq1}-\ref{eq2}) is:
\begin{equation}
\frac{\partial \bar{\mu}_1}{\partial
\gamma_2}=\frac{2\sqrt{1-\bar{\mu}_1}\, \bar{\mu}_1\,
(\bar{\sigma}^2-\bar{\mu}_1)(\bar{\sigma}^2-\bar{\mu}_2)}{(-1)^{\alpha_1}(\bar{\mu}_2-\bar{\mu}_1)}\,;\,
 \frac{\partial \bar{\mu}_2}{\partial
\gamma_2}=\frac{2\sqrt{1-\bar{\mu}_2}\, \bar{\mu}_2\,
(\bar{\sigma}^2-\bar{\mu}_2)(\bar{\sigma}^2-\bar{\mu}_2)}{(-1)^{\alpha_2}(\bar{\mu}_2-\bar{\mu}_1)}\label{jacobifield}
\end{equation}
The Jacobi field $V^1_J(x)=\frac{\partial \bar{\mu}_1}{\partial
\gamma_2}(x)$,  $V^2_J(x)=\frac{\partial \bar{\mu}_2}{\partial
\gamma_2}(x)$ is zero at the vacuum point. To check that it is
also zero at the focus we take the limit $\bar{\mu}_1\to
\bar{\sigma}^2$ from the left, and $\bar{\mu}_2\to \bar{\sigma}^2$
from the right:
\[
\lim_{(\bar{\mu}_1,\bar{\mu}_2)\to
(\bar{\sigma}^2,\bar{\sigma}^2)}\frac{\partial
\bar{\mu}_1}{\partial \gamma_2}  =
\lim_{(\bar{\mu}_1,\bar{\mu}_2)\to
(\bar{\sigma}^2,\bar{\sigma}^2)}
\frac{4\sigma^2\bar{\sigma}^2}{(-1)^{\alpha_1}}
\frac{1}{\frac{1}{\sqrt{1-\bar{\mu}_2}-\sigma}-\frac{1}{\sqrt{1-\bar{\mu}_1}-\sigma}}=0
\]
and the same thing happens to the $V^2_J$ component. Therefore,
the NTK3, NTK2$\sigma_3$ and NTK2$\sigma_2$ kinks are not local
minima of $E$: they are unstable.

\item {\bf TK3 topological kinks}

To identify the Jacobi fields on a given TK3 generic kink, we
shall perform an almost identical analysis and shall use equations
(\ref{69}) and (\ref{70}) to describe the TK3 orbit. We multiply
the equation for the TK3 kink form, member by member, by the
product of the $\sigma_3\bar{\sigma}_2^2$ power of (\ref{69}) by
the $-\sigma_2\bar{\sigma}_3^2$ power of (\ref{70}) to obtain the
modified kink form factor equation:
\begin{equation}
\left|\frac{\sqrt{1-\lambda_1}-1}{\sqrt{1-\lambda_1}+1}\right|^{(-1)^{\alpha_1}}\cdot
\left|\frac{\sqrt{1-\lambda_2}-1}{\sqrt{1-\lambda_2}+1}\right|^{(-1)^{\alpha_2}}\cdot
\left|\frac{\sqrt{1-\lambda_3}-1}{\sqrt{1-\lambda_3}+1}\right|^{(-1)^{\alpha_3}}=e^{2(x+\alpha)}\label{neweq0}
\end{equation}
where
$\alpha=\gamma_1+\frac{\bar{\sigma}_2^4}{\sigma_3^2-\sigma_2^2}\gamma-
\frac{\bar{\sigma}_3^4}{\sigma_3^2-\sigma_2^2}\bar{\gamma}$.

The vacuum point is a solution of (\ref{neweq0}) for $x=-\infty$,
$\alpha_1=\alpha_2=\alpha_3=0$ and $x=\infty$,
$\alpha_1=\alpha_2=\alpha_3=1$. Arguing along the same line as in
the NTK case, one finds that any TK3 trajectory hits the hyperbola
(\ref{34}) at the \lq\lq instant" $x_1=-\alpha-\frac{1}{2}\ln
\left|\frac{\sqrt{1-\tilde{\lambda}_1}-1}{\sqrt{1-\tilde{\lambda}_1}+1}
\right|^{(-1)^{\alpha_1}}$- $\tilde{\lambda}_1$ is a root of
equation (\ref{Iia}). The same trajectory reaches the ellipse
(\ref{33}) \lq\lq later ": $x_2=-\alpha-\frac{1}{2}\ln
\left|\frac{\sqrt{1-\tilde{\lambda}_3}-1}{\sqrt{1-\tilde{\lambda}_3}+1}
\right|^{(-1)^{\alpha_3}}$- $\tilde{\lambda}_3$ is a root of
equation (\ref{Ia}). The necessary fine tuning of the \lq\lq
arrival times" to the focal curves depends on the intersection
points of the trajectories with these curves. Thus, the focal
hyperbola and ellipse are indeed lines of conjugate points to the
vacuum.

The implicit derivation of (\ref{69}), (\ref{70}) and
(\ref{neweq0}) with respect to $\gamma$ and $\bar{\gamma}$, as
well as the solution of the subsequent algebraic linear system,
provides the two Jacobi fields orthogonal to each TK3 trajectory:
$\vec{V}_{J_1}(x)=\frac{\partial\vec{\lambda}_{{\rm
TK}3}}{\partial\gamma}(x)$,
$\vec{V}_{J_2}(x)=\frac{\partial\vec{\lambda}_{{\rm
TK}3}}{\partial\bar{\gamma}}(x)$. These fields are
\begin{eqnarray}
&&V^1_{J_1}(x)= -\frac{g^{11}}{\sigma_2^2-\sigma_3^2}
\frac{\partial W}{\partial \lambda_1}
(\bar{\lambda}_2-\bar{\sigma}_3^2)
(\bar{\lambda}_3-\bar{\sigma}_3^2);\  V^2_{J_1}(x)=
-\frac{g^{22}}{\sigma_2^2-\sigma_3^2} \frac{\partial W}{\partial
\lambda_2} (\bar{\lambda}_1-\bar{\sigma}_3^2)
(\bar{\lambda}_3-\bar{\sigma}_3^2)\nonumber\\ &&
V_{J_1}^3(x)=\frac{g^{33}}{\sigma_2^2-\sigma_3^2} \frac{\partial
W}{\partial \lambda_3} (\bar{\lambda}_1-\bar{\sigma}_3^2)
(\bar{\lambda}_2-\bar{\sigma}_3^2)\label{Jac2}
\\ &&
V_{J_2}^1(x)= -\frac{g^{11}}{\sigma_2^2-\sigma_3^2} \frac{\partial
W}{\partial \lambda_1} (\bar{\lambda}_2-\bar{\sigma}_2^2)
(\bar{\lambda}_3-\bar{\sigma}_2^2);\  V_{J_2}^2(x)=
-\frac{g^{22}}{\sigma_2^2-\sigma_3^2} \frac{\partial W}{\partial
\lambda_2} (\bar{\lambda}_1-\bar{\sigma}_2^2)
(\bar{\lambda}_3-\bar{\sigma}_2^2)\nonumber \\ &&V_{J_2}^3(x)=
-\frac{g^{33}}{\sigma_2^2-\sigma_3^2} \frac{\partial W}{\partial
\lambda_3} (\bar{\lambda}_1-\bar{\sigma}_2^2)
(\bar{\lambda}_2-\bar{\sigma}_2^2)\label{Jac1}
\end{eqnarray}
We have that:
\[
\lim_{(\bar{\lambda}_1,\bar{\lambda}_2,\bar{\lambda}_3)\to
(\tilde{\lambda}_1,\bar{\sigma}_2^2,\bar{\sigma}_2^2)} V^1_{J_1}=0
\quad , \quad
\lim_{(\bar{\lambda}_1,\bar{\lambda}_2,\bar{\lambda}_3)\to
(\tilde{\lambda}_1,\bar{\sigma}_2^2,\bar{\sigma}_2^2)}
V^2_{J_1}=\frac{0}{0}\quad , \quad
\lim_{(\bar{\lambda}_1,\bar{\lambda}_2,\bar{\lambda}_3)\to
(\tilde{\lambda}_1,\bar{\sigma}_2^2,\bar{\sigma}_2^2)}
V^3_{J_1}=\frac{0}{0}
\]
and the $V_{J_1}^1$ component of the Jacobi field is zero on the
focal hyperbola, whereas $V_{J_1}^2, V_{J_1}^3$ are indeterminate.
To solve for the indeterminacy in the second component, we
carefully take the limit of $V^2_{J_1}$ when $\bar{\lambda}_2$ and
$\bar{\lambda}_3$ go to $\bar{\sigma}_2^2$ respectively from the
left and from the right:
\[
\lim_{(\bar{\lambda}_1,\bar{\lambda}_2,\bar{\lambda}_3)\to
(\tilde{\lambda}_1,\bar{\sigma}_2^2,\bar{\sigma}_2^2)}
V_{J_1}^2=\frac{4\sigma_2^2\bar{\sigma}_2^2
(\tilde{\lambda}_1-\bar{\sigma}_3^2)}{(\tilde{\lambda}_1-\bar{\sigma}_2^2)(-1)^{\alpha_1}}
\lim_{(\bar{\lambda}_1,\bar{\lambda}_2,\bar{\lambda}_3)\to
(\tilde{\lambda}_1,\bar{\sigma}_2^2,\bar{\sigma}_2^2)}\frac{1}{\frac{1}{\sqrt{1-\bar{\lambda}_2}
-\sigma_2}-\frac{1}{\sqrt{1-\bar{\lambda}_3} -\sigma_2}}=0
\]
The same limit shows that the third component is also zero on the
focal hyperbola and that the Jacobi field $\vec{V}_{J_1}(x)$ is
zero at the intersection points of the TK3 trajectory with the
hyperbola (\ref{Iia}). The zeroes of the other Jacobi field,
$\vec{V}_{J_2}(x)$, are the vacuum point and the intersection
point of the kink trajectory with the ellipse (\ref{Ia}). The TK3
kinks are saddle points of the energy functional and therefore are
unstable.

\end{enumerate}

\subsection{The spectrum of the Hessian for the non-generic topological kinks}

Non-generic topological kinks are very singular points at the
boundary of the kink moduli space and the extension of this
procedure to them could cast doubt on its applicability.
Fortunately, analytical formulas both for the kink orbit and form
factors are available in Cartesian coordinates and one can
directly look at the spectrum of the Hessian:
\[
\Delta\vec{V}=\sum_{a=1}^3\sum_{b=1}^3\Delta^a_bV^b\vec{e}_a\qquad.
\]

\subsubsection{TK1 Hessian}
In this case the Hessian is a diagonal matrix, $\Delta^a_b=0,
a\neq b$, of Schrodinger operators
\begin{equation}
\Delta^1_1 =-\frac{d^2}{dx^2}+4-\frac{6}{\cosh^2x}\quad , \quad
\Delta^2_2= -\frac{d^2}{dx^2}+\sigma_2^2-\frac{2}{\cosh^2x} \quad
, \quad \Delta^3_3=
-\frac{d^2}{dx^2}+\sigma_3^2-\frac{2}{\cosh^2x}\label{hessianTK1}
\end{equation}
with potential wells of P\"osch-Teller type. The spectrum is shown
in the next Table:
\begin{center}
\begin{tabular}{|c|c|c|} \hline
  $\Delta^1_1$ & $\Delta^2_2$ & $\Delta^3_3$ \\[0.2cm] \hline
 $(0)\sqcup (3)\sqcup (k_1^2+4)$ & $(-\bar{\sigma}_2^2)\sqcup (0)
 \sqcup (k_2^2+\sigma_2^2)$ & $(-\bar{\sigma}_3^2)\sqcup (0)
 \sqcup (k_3^2+\sigma_3^2)$
 \\[0.2cm] \hline
 $\vec{V}_{(0)}(x)={\rm sech}^2x\vec{e}_1$ & $\vec{V}_{(-\bar{\sigma}_2^2)}(x)
 ={\rm sech}x\vec{e}_2$ & $\vec{V}_{(-\bar{\sigma}_3^2)}(x)
 ={\rm sech}x\vec{e}_3$  \\[0.2cm] \hline $\vec{V}_{(3)}(x)=
\frac{{\rm tanh}x}{{\rm cosh}x}\vec{e}_1$ &
$\vec{V}_{J_2}(x)=e^{\sigma_2x}(\sigma_2-{\rm tanh}x)\vec{e}_2$ &
$\vec{V}_{J_3}(x)=e^{\sigma_3x}(\sigma_3-{\rm tanh}x)\vec{e}_3$
\\ [0.2cm] \hline $\vec{V}_{(k_1^2)}(x)=e^{ik_1x}P_2({\rm
tanh}x)\vec{e}_1$ & $\vec{V}_{(k_2^2)}(x)=e^{ik_2x}P_1({\rm
tanh}x)\vec{e}_2$ & $\vec{V}_{(k_3^2)}(x)=e^{ik_3x}P_1({\rm
tanh}x)\vec{e}_3$
\\[0.2cm]\hline
\end{tabular}
\end{center}
Here, $k_1,k_2,k_3\in{\Bbb R}$ and $P_l({\rm tanh}x)$ is the
$l^{{\rm th}}$ Jacobi polynomial. The spectrum of $\Delta$
contains two negative eigenvalues obeying small fluctuations in
the directions  orthogonal to the kink orbit. The TK1 kink is a
saddle point of $E[\vec{\phi}]$ and is therefore unstable. The
Jacobi fields $\vec{V}_{J_1}$ and $\vec{V}_{J_2}$ are both zero at
$x=-\infty$ when the trajectory departs from the vacuum point
$\vec{v}=-\vec{e}_1$, and reach a second zero respectively at
$x={\rm arctanh}\sigma_2$, $x={\rm arctanh}\sigma_3$, the points
where the foci F$_2\equiv(\sigma_2,0,0)$,
F$_3\equiv(\sigma_3,0,0)$ are hit by the TK1 trajectory, as
expected.

\subsubsection{TK2$\sigma_3$ and TK2$\sigma_2$ Hessians}
The Hessian for the TK2$\sigma^3$ kink is non-diagonal:
\[
\Delta \vec{V}\equiv  \left( \begin{array}{ccc}
-\frac{d^2}{dx^2}+4-\frac{2(2+\sigma_3^2)}{\cosh^2(\sigma_3 x)}
&0& 4 \bar{\sigma_3} \frac{\tanh (\sigma_3 x)}{\cosh (\sigma_3 x)}
\\
0&-\frac{d^2}{dx^2}+\sigma_2^2-\frac{2\sigma_3^2}{\cosh^2(\sigma_3
x)}&0\\  4\bar{\sigma_3} \frac{\tanh (\sigma_3 x)}{\cosh (\sigma_3
x)}&0&
-\frac{d^2}{dx^2}+\sigma_3^2-\frac{2(3\sigma_3^2-2)}{\cosh^2(\sigma_3
x)}
\end{array} \right) \left( \begin{array}{c} V^1 \\ V^2 \\ V^3
\end{array}\right)
\]
We simply study the spectrum of $\Delta^2_2$, governing the small
fluctuations in the direction  orthogonal to the
$\vec{e}_1;\vec{e}_3$ plane, where the TK2$\sigma_3$ kink lives.
The first eigenvalue, $\sigma^2_2-\sigma^2_3$, with an
eigen-function
$\vec{V}_{(\sigma_2^2-\sigma_3^2)}(x)=\frac{1}{\cosh (\sigma_3
x)}\vec{e}_2$, shows that the TK2$\sigma_3$ kink is a saddle point
of $E[\vec{\phi}]$; it is, therefore, unstable. Moreover, there is
a Jacobi field $\vec{V}_{J}(x)=e^{\sigma_2 x}\left(
\frac{\sigma_2}{\sigma_3}-\tanh (\sigma_3 x)\right)\vec{e}_2$,
which, besides the vacuum point, is zero at the umbilicus: $x=
\frac{1}{\sigma_3}{\rm arctanh}\frac{\sigma_2}{\sigma_3}$ yields
$\phi^1=\frac{\sigma_2}{\sigma_3}$, which is the umbilicus point
of the ellipsoid on the TK2$\sigma_3$ orbit. After crossing the
umbilicus, this field can be glued in a continuous but not
differentiable way to the $\vec{V}_{J_2}$ Jacobi field on the TK1
kink beyond the focus $F_2$. The new field can be understood as a
Jacobi field associated with the $T_2^{\sigma_2}T_2^{\sigma_3}T_1$
limit of the TK3 moduli space. Note that there are also
eigen-functions in the continuous spectrum with eigenvalues
$k^2\sigma_3^2+\sigma_2^2$.

The Hessian for the TK2$\sigma_2$ kink is almost identical:
\[
\Delta \vec{V}\equiv  \left( \begin{array}{ccc}
-\frac{d^2}{dx^2}+4-\frac{2(2+\sigma_2^2)}{\cosh^2(\sigma_2 x)} &4
\bar{\sigma_2} \frac{\tanh (\sigma_2 x)}{\cosh (\sigma_2 x)}& 0
\\
4\bar{\sigma_2} \frac{\tanh (\sigma_2 x)}{\cosh (\sigma_2
x)}&-\frac{d^2}{dx^2}+\sigma_2^2-\frac{2(3\sigma_2^2-2)}{\cosh^2(\sigma_2
x)}&0\\ 0&0&
-\frac{d^2}{dx^2}+\sigma_3^2-\frac{2\sigma_2^2}{\cosh^2(\sigma_2
x)}
\end{array} \right) \left( \begin{array}{c} V^1 \\ V^2 \\ V^3
\end{array}\right)
\]
The non-diagonal box, however, governs small fluctuations on the
$\vec{e}_1;\vec{e}_2$ plane, instead of on the
$\vec{e}_1;\vec{e}_3$ plane. The most important difference is that
the Jacobi field $\vec{V}_{J}(x)=e^{\sigma_3 x}\left(
\frac{\sigma_3}{\sigma_2}-\tanh (\sigma_2 x)\right)\vec{e}_3$ of
$\Delta_3^3$ is only zero at the vacuum point on the TK2$\sigma_2$
trajectory. This field intersects the Jacobi field $\vec{V}_{J_3}$
on the TK1 kink -after this latter has crossed the focus $F_3$ -
and both form a new Jacobi field on the
T$_2^{\sigma_2}$T$_2^{\sigma_3}$T$_1$ kink configuration . The
spectrum of $\Delta_3^3$ starts with the positive eigenvalue
$\sigma_3^2-\sigma_2^2$ and the TK2$\sigma_2$ kink is stable
against small fluctuations that bring the orbit away from the
$\vec{e}_1;\vec{e}_2$ plane.

Concerning fluctuations of the form
$\vec{V}(x)=V^1(x)\vec{e}_1+V^2(x)\vec{e}_2$, we cannot explicitly
compute the spectrum of $\Delta$: one should find the stationary
states for a quantum $\frac{1}{2}$ spin particle moving through a
non-constant spin-exchange potential. Nevertheless, because
TK2$\sigma_2$ is an absolute minimum of the energy, there cannot
be negative eigenvalues in Spec$\Delta$. In fact, we know the
translational mode: $\vec{V}_{(0)}(x)=\frac{1}{{\rm
cosh}(\sigma_2x)}(\vec{e}_1-\bar{\sigma}_2{\rm
sinh}(\sigma_2x)\vec{e}_2)$, which is tangent to the TK2$\sigma_2$
orbit. There must also be a Jacobi field, null only at the vacuum
point, along the TK2$\sigma_2$ orbit. This Jacobi field is
completed by $\vec{V}_{J_2}(x)$ after their intersection, which
occurs before the crossing of $\vec{V}_{J_2}(x)$ by $F_2$.

\subsection{Morse theory of kinks}
Let $\Omega M^n=\{ \gamma:S^1\to M^n\, /\,
\gamma(0)=\gamma(1)=m_0\}$ be the space of closed paths with a
fixed base point $m_0$ in a Riemannian manifold $M^n$. Morse
theory establishes a link between the topology of $\Omega M^n$ and
the \lq\lq critical point" structure of a well-defined functional
in $\Omega M^n$, \cite{11}. A formulation of this theory \`a la
Bott, \cite{12}, is as follows:

Let $P_t\left( \Omega M^n \right)=\sum_{k=0}^{\infty} b_k\, t^k$
be the Poincar\`e series of $\Omega M^n$ in the indeterminate $t$.
The topology of $\Omega M^n$ is encoded in $P_t\left( \Omega
M^n\right)$ because $b_k={\rm dim} \, H_k\left( \Omega M^n, {\Bbb
R}\right)$ are the Betti numbers that account for the dimension of
the $k^{\rm th}$ homology group of $\Omega M^n$. Let ${\cal
M}_t\left( E\right) =\sum_{N_c}\, P_t\left( N_c\right)\,
t^{\mu(N_c)}$ be the Morse series of the $E$ functional. ${\cal
M}_t(E)$ encodes the critical point structure of $E$: the sum is
over the critical manifolds $N_c$ formed either by a single
isolated critical path or by continuous sets of critical paths,
depending on $M^n$. $P_t(N_c)$ is the Poincar\'e polynomial of
$N_c$ and the most important ingredient is the Morse index
$\mu(N_c)$:

$\mu(N_c)$ is the dimension of the sub-space in the tangent space
$T_{\gamma_c}\Omega M^n$ where the Hessian quadratic form of $L$
at $\gamma_c\in N_c$ is negative definite in the directions
orthogonal to $N_c$. The Morse inequalities,
\[
{\cal M}_t(E)\, \geq\, P_t\left( \Omega M^n\right)
\]
tell us that the topology of $\Omega M^n$ forces the existence of
most of the critical points of $E$.

The Morse index can also be understood as the number of negative
eigenvalues of the Hessian operator and hence $\mu(N_c)$ informs
us about the (lack of) stability of a given critical path
$\gamma_c$. On the other hand, one can skip the solving of a
difficult spectral problem to compute $\mu(N_c)$ by applying the
\underline{Morse index theorem}:

\lq\lq The Morse index of a critical path $\gamma_c$ is equal to
the number of conjugate points to the base point crossed by
$\gamma_c$ counted with multiplicity". Recall that conjugate
points to $m_0$ are those points in $M^n$ where infinite critical
paths passing through $m_0$ meet again and the multiplicity is the
dimension of a section transverse to this congruence near the
focal point, \cite{8} and \cite{9}.

In this sub-Section we shall develop the Morse theory of kinks in
our system as a synthetic approach for dealing with the stability
problem. We have the following dictionary: $M^n$ is the manifold
$\bar{\bf P}_3(0)$; $\Omega M^n$ is the full configuration space
${\cal C}$. Note that in elliptic coordinates the four
disconnected sectors are melded in a single sector: ${\cal
C}={\cal C}^{11}\sqcup {\cal C}^{12} \sqcup {\cal C}^{21} \sqcup
{\cal C}^{22}.$ A topology in ${\cal C}$ is defined in Reference
\cite{9}. The real homology of ${\cal C}$ is also analysed in
\cite{9}. $m_0$ is the vacuum point $D$ in $\bar{\bf P}_3(0)$.

The conjugate points to D have been determined in the previous
sub-sections \S 4.1 and \S 4.2 as zero loci of the Jacobi fields.
We summarize the results as follows:

\begin{itemize}

\item (1). If $\lambda_1=\bar{\sigma}_3^2=\lambda_2$. Every point in
the F$_1$F$_3$ edge is a conjugate point to D. Through any point
in the focal line, F$_1$F$_3$ crosses a one-parameter congruence
of kink trajectories, the multiplicity of the conjugate points in
F$_1$F$_3$ is 1.

\item (2). If $\lambda_2=\bar{\sigma}_2^2=\lambda_3$. Each point in the
AF$_2$ edge is also a conjugate point to D. AF$_2$ is a focal line
formed by conjugate points of multiplicity 1.

\end{itemize}

The points A, F$_2$ and F$_3$ are also conjugate points. In fact,
these points are crossed by the kinks belonging to two
one-dimensional families. The trajectories in the kink moduli
spaces ${\cal M}_{{\rm N}_3{\rm T}_1}$ and ${\cal M}_{{\rm
N}_2^{\sigma_3}{\rm T}_2^{\sigma_2}}$ pass through the F$_3$
focus; either the N$_3$T$_1$ or the
N$_2^{\sigma_3}$T$_2^{\sigma_2}$ family is reached, depending on
which path in ${\cal M}_{{\rm T}_3}$ is chosen. Simili modo, all
the kink trajectories in both ${\cal M}_{{\rm N}_3{\rm T}_1}$ and
${\cal M}_{{\rm N}_2^{\sigma_2}{\rm T}_2^{\sigma_3}}$ coincide at
the umbilicus A and the F$_2$ focus; again, the two possibilities
arise in connection with the existence of two paths in ${\cal
M}_{{\rm T}_3}$ that lead to different families of singular kinks.

Finally D itself can be viewed as a conjugate point of
multiplicity 2; there is a two-parameter congruence of kink
trajectories emanating from D.

\subsubsection{The Morse series of ${\cal C}$}

We now apply the Morse index theorem to compute the Morse index of
the kink trajectories.

\medskip

\noindent {\bf I.} Vacuum trajectory: D. Morse index, $\mu({\rm
D})=0$. Here, we include the computation of the TK2$\sigma_2$ kink
as the only stable configuration, and therefore consider the edge
DC deformed continuously to the vacuum point D.

\medskip

\noindent {\bf II.} TK2$\sigma_3 \, \sqcup$ AK2$\sigma_3$.

The TK2$\sigma_3$ trajectory starts from D, passes through the
umbilicus point A, reaches the point B, and comes back to D along
the same path. A subtle point is that only in the second crossing
is the A point a conjugate point to D along the TK2$\sigma_3$
trajectory because the last crossing happens at the same time as a
congruence of N$_3$T$_1$ kinks reaches point A in the edge AF$_2$
in its first crossing. Another even more subtle point is the
following: we are considering closed trajectories, but the
TK2$\sigma_3$ trajectory is closed only in $\bar{\bf P}_3(0)$. If
we require trajectories also closed in Euclidean space, the
TK2$\sigma_3$ trajectory must be counted together with the
anti-kink AK2$\sigma_3$. Again, only in the second crossing along
the AK2$\sigma_3$ trajectory is the umbilicus point A a conjugate
point to D. Therefore, the Morse index of the TK2$\sigma_3\,
\sqcup$ AK2$\sigma_3$ trajectory is two: $\mu\left( {\rm
TK2}\sigma_3\, \sqcup\, {\rm AK2}\sigma_3\right)=2$. Note that the
second crossing of the TK3 and AK3 trajectories though A are
simultaneous and that D is only the starting and ending point of
this super-imposed kink.

\medskip

\noindent {\bf III.} TK1 $\sqcup$ AK1.

Both the TK1 and AK1 trajectories pass through the conjugate
points to D F$_2$ and F$_3$ twice, but only in the second crossing
are they truly conjugate points to D because then, these points
are reached by the TK3 or AK3 congruences for the first time.
Therefore, the Morse index of the TK1 $\sqcup$ AK1 trajectory is
four: $\mu\left( {\rm TK1} \sqcup {\rm AK1}\right)=4$.

\medskip

\noindent {\bf IV.} TK3 $\sqcup$ AK3.

A TK3 trajectory hits the F$_1$F$_3$ edge once and the AF$_2$ edge
twice, whereas the anti-kink AK3 trajectory does the same in the
opposite sense. Therefore, $\mu({\rm TK3}\sqcup {\rm AK3})=6$. The
critical manifold formed by these configurations is: ${\cal
N}_c=\bar{\cal M}_{T_3}^{(3)}\sqcup\bar{\cal M}_{A_3}^{(3)}$. We
have chosen the compactification of ${\cal M}_{T_3}$ described in
sub-Section \S 3.3: it is understood that in the count of a
trajectory passing through A, F$_2$ or F$_3$ the appropriate
N$_3$T$_1$, N$_2^{\sigma_3}$T$_2^{\sigma_2}$ or
N$_2^{\sigma_3}$T$_2^{\sigma_2}$ congruence is taken into account.
In any case, $P_t(\bar{{\cal M}}_{{\rm T}_3}^{(3)}\sqcup
\bar{{\cal M}}_{{\rm A}_3}^{(3)})=1+t^2$ and the contribution to
the Morse series is $(1+ t^2) t^6$.

In fact, these are the only kink trajectories that should be
included, because all of them starts and end at D. Strictly
speaking, D cannot be crossed by any trajectory because D only can
be reached again in infinite \lq\lq time".

Closing our eyes to this fact, the next critical manifold is:

\medskip

\noindent {\bf V.} (TK2$\sigma_3 \, \#$ TK3(1)) $\sqcup$
(AK2$\sigma_3 \,  \#$ AK3(1)).

The TK2$\sigma_3$ trajectory is followed by a single TK3(1) kink:
precisely T$_2^{\sigma_2}$T$_2^{\sigma_3}$T$_1$ . The two
trajectories are glued (represented by the symbol $\#$) at point
D. The crossing of D, which is shared by the composite kink and
the anti-kink trajectories, contributes to the Morse index with 2
because of the multiplicity:
\[
\mu\left( ({\rm T}_2^{\sigma_3} \, \# {\rm T}_2^{\sigma_2}{\rm
T}_2^{\sigma_3}{\rm T}_1) \sqcup ({\rm A}_2^{\sigma_3} \, \# {\rm
A}_2^{\sigma_2}{\rm A}_2^{\sigma_3}{\rm A}_1) \right) =10
\]

\medskip

\noindent {\bf VI.} (TK1 $ \#$ TK3(1)) $\sqcup$ (AK1 $ \#$
AK3(1)).

For the same reasons,
\[
\mu\left( ({\rm T}_1 \, \# {\rm T}_2^{\sigma_2}{\rm
T}_2^{\sigma_3}{\rm T}_1) \sqcup ({\rm A}_1\, \# {\rm
A}_2^{\sigma_2}{\rm A}_2^{\sigma_3}{\rm A}_1) \right) =12
\]

\medskip

\noindent {\bf VII.} (TK3 $\#$ TK3) $\sqcup$ (AK3 $\#$ AK3).

The gluing of two families of TK3 and AK3 kinks again requires the
application of degenerate Morse theory. The Morse index of one
member of the family is $\mu( ({\rm TK3} \# {\rm TK3}) \sqcup (
{\rm AK3} \# {\rm AK3}))=14$ but the Poincar\`e polynomial of this
critical manifold is: $P_t(({\bar{\cal M}}_{{\rm T}_3} \#
\bar{{\cal M}}_{{\rm T}_3}) \sqcup ( \bar{{\cal M}}_{{\rm A}_3} \#
\bar{{\cal M}}_{{\rm A}_3}))=1+t^2$.

\medskip

Iteration of these basic elements produces the Morse series:
\begin{eqnarray*}
{\cal M}_t\left(\left. E\right|_{\cal C}\right) &=& 1+ \left(
1+t^2\right) \, t^2+ \left( 1+t^2\right) \, t^6+\left(
1+t^2\right) \, t^{10} + \left( 1+t^2\right) \, t^{14}+\dots \\
&=& 1+\left( 1+t^2\right) \, \sum_{k=0}^\infty \, t^{2(2k+1)}\,
=\, \frac{1}{1-t^2}
\end{eqnarray*}

Because all the coefficients of the odd powers of $t$ in ${\cal
M}_t$ are zero, the lacunary principle states that the Morse
inequality becomes equality, \cite{12}:
\[
{\cal M}_t\left(\left. E\right|_{\cal C}\right)=P_t\left( {\cal
C}\right) =\frac{1}{1-t^2}= P_t\left( \Omega S^3\right)
\]

The configuration space is of the same homology type as $\Omega
S^3$: the loop space in $S^3$. There is some kind of topological
universality and the first terms in ${\cal M}_t\left(\left.
E\right|_{\cal C}\right)$ tell us that the existence of
TK2$\sigma_3 \, \sqcup$ AK2$\sigma_3$ kinks is due to the fact
that the homology group $H_2({\cal C},{\Bbb R})={\Bbb R}$ is
non-trivial; the TK1 $\sqcup$ AK1 kink comes from $H_4({\cal
C},{\Bbb R})={\Bbb R}$; the TK3 $\sqcup$ AK3 family corresponds to
$H_6({\cal C},{\Bbb R})={\Bbb R}$, and so forth.

\subsubsection{Kink stability from the index theorem}
We have shown in sub-sections \S 4.1 and \S 4.2 that in the
topological sectors ${\cal C}^{12}$ and  ${\cal C}^{21}$ only the
TK2$\sigma_2$ and AK2$\sigma_2$ kinks are stable, whereas in the
non-topological sectors ${\cal C}^{11}$ and  ${\cal C}^{22}$ only
the ground states $v^1$ and $v^2$ are stable. The degree of
instability of each kind of kink is measured by the Morse index,
which agrees with the number of kink configurations with lower
energy in the same topological sector. We have a situation of
stratified Morse theory according to the energy hierarchy in each
topological sector:

\medskip

\noindent A. ${\cal C}^{12}$ and ${\cal C}^{21}$:
\begin{eqnarray*}
&& E({\rm TK3})>E({\rm TK1})>E({\rm TK2}\sigma_3)>E({\rm
TK2}\sigma_2)\\ && E({\rm AK3})>E({\rm AK1})>E({\rm
AK2}\sigma_3)>E({\rm AK2}\sigma_2)
\end{eqnarray*}

\noindent B. ${\cal C}^{11}$ and ${\cal C}^{22}$: $\alpha=1,2$
\[
E({\rm NTK2}\sigma_3(\alpha))>E({\rm NTK2}\sigma_2(\alpha))>E({\rm
NTK3}(\alpha))>E(v^{\alpha})
\]
where NTK$(\alpha)$ means that the NTK family starts from the
vacuum $v^{\alpha}$.

To check this statement, we simply apply the Morse index theorem
to compute the Morse index of each kink in the system .
\medskip

\noindent {\bf 1.} Kinks in the topological sector ${\cal
C}^{12}$:

\begin{description}

\item{A.} $\mu({\rm TK3})=3$. The application of the Morse index theorem to a
TK3 trajectory works as follows: the trajectory starts at the
ground state $v^1$, goes to the focal hyperbola
$\frac{\phi_1^2}{\sigma_2^2}-\frac{\phi_3^2}{\sigma_3^2-\sigma_2^2}=1$,
comes back to the focal ellipse
$\frac{\phi_1^2}{\sigma_3^2}+\frac{\phi_2^2}{\sigma_3^2-\sigma_2^2}=1$,
reaches the focal hyperbola (the other branch) again, and ends at
$v^2$. Therefore, a TK3 trajectory crosses three c. p. to D, each
of multiplicity one. There are three orthogonal directions of
instability in ${\cal C}$ for any TK3 kink.

\item{B.} $\mu({\rm TK1})=2$. It is easy to repeat in Cartesian coordinates
the arguments presented in elliptic coordinates above, in order to
conclude that the Morse index of a TK1 kink is 2. The TK1
trajectory crosses the foci F$_2$ and F$_3$ of the ellipsoid
$\phi_1^2+\frac{\phi_2^2}{\bar{\sigma}_2^2}+\frac{\phi_3^2}{\bar{\sigma}_3^2}=1$
twice, but only in the second crossing are the foci reached by the
NTK2$\sigma_2$ or NTK2$\sigma_3$ trajectories. There are only two
directions of instability near the TK1 kink, as has been
explicitly shown in sub-section \S 4.2.2.

\item{C.} $\mu({\rm TK2}\sigma_3)=1$. A TK2$\sigma_3$ trajectory crosses
two umbilicus points in the ellipsoid, but only in the second
crossing is the umbilicus point a conjugate point to $v^1$ along
the TK2$\sigma_3$ path because all the NTK3 congruence reaches the
second umbilicus at the same instant as TK2$\sigma_3$. The
direction of instability in ${\cal C}$ near TK2$\sigma_3$ has also
been shown in sub-section \S 4.2.2.

\item{D.} $\mu({\rm TK2}\sigma_2)=0$. The TK2$\sigma_2$ trajectories do
not cross any focal lines and their Morse index is zero. Again, in
sub-section \S 4.2.2 we saw that there are no directions of
instability near TK2$\sigma_2$ by showing that the Hessian
quadratic form of $E$ at TK2$\sigma_2$ is semi-definite positive.

\end{description}

\noindent {\bf 2.} Kinks in the non-topological sector ${\cal
C}^{11}$:

\begin{description}

\item{A.} $\mu({\rm NTK2}\sigma_3)=3$. Any NTK2$\sigma_3$ trajectory hits the focal
hyperbola
$\frac{\phi_1^2}{\sigma_2^2}-\frac{\phi_3^2}{\sigma_3^2-\sigma_2^2}=1$
four times. Only during the second and fourth crossings does the
NTK2$\sigma_3$ trajectory reach conjugate points to $v^1$.

The time table is as follows: 1) the second crossing of the
NTK2$\sigma_3$ trajectory by the focal hyperbola coincides with
the first crossing of a TK3 congruence. 2) The NTK2$\sigma_3$ kink
crosses the F$_3$ focus at the same time as any other
NTK2$\sigma_3$. 3) The fourth crossing of the NTK2$\sigma_3$
trajectory by the focal hyperbola (second branch) happens
simultaneously to the crossing of the TK3 congruence. Therefore,
the Morse index of the NTK2$\sigma_3$ is 3: there are three
directions of instability in $T_{{\rm NTK2}\sigma_3}{\cal C}$.

\item{B.} $\mu({\rm NTK2}\sigma_2)=2$. Any NTK2$\sigma_2$ trajectory
hits the focal ellipse
$\frac{\phi_1^2}{\sigma_3^2}+\frac{\phi_2^2}{\sigma_3^2-\sigma_2^2}=1$
twice. Only during the second crossing does the NTK2$\sigma_2$
trajectory reach a conjugate point to $v^1$. The time table is as
follows: 1) All the members of the NTK2$\sigma_2$ congruence meet
at the F$_2$ focus at the same time. F$_2$ is thus a conjugate
point to $v^1$ with multiplicity one. 2) The second crossing of
the NTK2$\sigma_2$ kinks through the focal ellipse is simultaneous
to the crossing of a TK3 congruence; it is the second conjugate
point of multiplicity one to $v^1$ along any NTK2$\sigma_2$ kink.
The first crossing of the focal ellipse, however, does not
correspond to a conjugate point to $v^2$. The Morse index is 2:
there are two orthogonal instability directions in $T_{{\rm
NTK2}\sigma_2}{\cal C}$.

\item{C.} $\mu({\rm NTK3})=1$. Any member of the NTK3 family intersects
at an umbilicus point at the same time. The umbilicus is the only
conjugate point, with multiplicity one, to $v^1$ along the NTK3
trajectory and the NTK3 Morse index is one.

\item{D.} $\mu(v^1)=0$. The only stable trajectory in ${\cal C}^{11}$ is the ground state $v^1$.

\end{description}

In sum, there is a stratification of each connected component
${\cal C}^{\alpha\beta}$ of the configuration space ${\cal C}$ by
the \lq\lq critical" points of the functional $E$. Morse's theory
relates the structure of such critical points with the homology of
the configuration space itself.

\section{Further Comments }

The stability characteristics of classical solutions in field
theory provide qualitative insight about the behaviour of the
quantum kink descendants in the semi-classical limit that usually
survives strong quantum fluctuations. Therefore, there is a bona
fide quantum ${\rm TK}2\sigma_2$ kink state which belongs to the
spectrum of the quantum Hamiltonian of the system. The energy up
to one-loop order is:
\[
\hat{E}[\vec{\Phi}_{T_2^{\sigma_2}}]={E}[\vec{\Phi}_{T_2^{\sigma_2}}]+\frac{\hbar}{2}c_d\left({\rm
Tr}K^{\frac{1}{2}}(\vec{\Phi}_{T_2^{\sigma_2}})-{\rm
Tr}K^{\frac{1}{2}}(\vec{\Phi}_V)+\Delta_m(\vec{\Phi}_{T_2^{\sigma_2}})-
\Delta_m(\vec{\Phi}_V)\right)+O(\hbar^2)
\]
Here,
\[
{\rm
K}_{ab}(\vec{\Phi}_C)=-\frac{d^2}{dx^2}+\frac{\delta^2U}{\delta\phi^a\delta\phi^b}|_{\vec{\Phi}_C},
\]
$c_d$ is a dimension-full parameter, and
\[
\Delta_m(\vec{\Phi}_C)=\sum_{a=1}^3\delta m_{aa}.\int dx
\frac{\delta^2U}{\delta\phi^a\delta\phi^a}|_{\vec{\Phi}_C}
\]
are the counter-terms induced by mass renormalization to take care
of the infinite $\delta m_{aa}$ contributions coming from the
tadpole graphs.

All the other kink solutions give rise to resonant states because
they are \lq\lq sphalerons" rather than solitons and decay either
to the vacuum or the ${\rm TK}2\sigma_2$ kink, as explained in
Reference \cite{15}.

\appendix
\section*{Appendix: The super-potential in Cartesian coordinates}
\addcontentsline{toc}{section}{Appendix: The super-potential in
Cartesian coordinates}

Solving for $\lambda_1$, $\lambda_2$, $\lambda_3$ in (\ref{17}) as
functions of $\phi_1$, $\phi_2$, $\phi_3$ demands that one must
solve:
\begin{equation}
\sum_{a=1}^3 \frac{\phi_a^2}{{\bar\sigma}_a^2-\lambda}=1
,\label{290}
\end{equation}
(see Appendix in Reference \cite{1}) which can be written as a
cubic algebraic equation in the complex variable $\lambda \in
{\Bbb C}$. Cardano`s formulae provides the roots of the cubic in
the form:
\begin{eqnarray*}
\lambda_1&=& -\sqrt{| q|} \left( \cos \frac{\theta}{3}+\sqrt{3}
\sin \frac{\theta}{3}\right)-\frac{u-A_1}{3};\hspace{0.8cm}
\lambda_2= -\sqrt{| q|} \left( \cos \frac{\theta}{3}-\sqrt{3} \sin
\frac{\theta}{3}\right)-\frac{u-A_1}{3}\\\ \lambda_3&=& 2 \sqrt{|
q|}  \cos \frac{\theta}{3}-\frac{u-A_1}{3}
\end{eqnarray*}
where
\[
|q|=-q=\frac{1}{9}\left( \left(
\phi_1^2+\phi_2^2+\phi_3^2\right)^2-\left(\sigma_2^2+\sigma_3^2\right)
\phi_1^2+\left( 2 \sigma_2^2-\sigma_3^2\right) \phi_2^2+\left( 2
\sigma_3^2-\sigma_2^2\right)
\phi_3^2+\sigma_2^4+\sigma_3^4-\sigma_2^2 \sigma_3^2\right)
\]
\begin{eqnarray*}
r&=&\frac{-1}{54} \left(
2(\sigma_2^6+\sigma_3^6)-3(\sigma_2^4\sigma_3^2+\sigma_2^2
\sigma_3^4)\right) +\frac{1}{18} \left(  \sigma_2^4 +\sigma_3^4
-4\sigma_2^2\sigma_3^2\right)\cdot\left(
\phi_1^2+\phi_2^2+\phi_3^2\right)+\\ && +\frac{1}{18}
(\sigma_2^2+\sigma_3^2)\left( \phi_1^2+\phi_2^2+\phi_3^2\right)^2
-\frac{1}{6} \left( \phi_1^2+\phi_2^2+\phi_3^2\right) \left(
\sigma_2^2 \phi_2^2+\sigma_3^2 \phi_3^2\right)+\\ &&+\frac{1}{6}
\left( 2\sigma_2^2\sigma_3^2-\sigma_2^4 \right)
\phi_2^2+\frac{1}{6}\left( 2\sigma_2^2\sigma_3^2-\sigma_3^4\right)
\phi_3^2-\frac{1}{27} \left( \phi_1^2+\phi_2^2+\phi_3^2\right)^3
\end{eqnarray*}
and the angle $\theta=\arctan \frac{\sqrt{|q^3+r^2|}}{r}$ is
defined in the open interval: $\theta \in (0,\pi)$. Note that: (a)
$ q^3+r^2$ is semi-definite negative; $q^3+r^2\neq 0$ implies
three distinct real roots. (b) $\theta =0$ and $\theta=\pi$
correspond to $|q^3+r^2|=0$ and hence these are values where the
square root is ill-defined. (c) choosing another open interval,
$\theta \in (\pi,2\pi)$ for instance, leads to a permutation of
the $\lambda_a$ roots due to the ${\Bbb Z}_3$ symmetry.

Using the previous results we write the super-potential in
Cartesian coordinates:
\begin{eqnarray}
&&W^{(\alpha_1,\alpha_2,\alpha_3 )}(\phi_1,\phi_2,\phi_3)=
\frac{1}{3\sqrt{3}} \left[ (-1)^{\alpha_1} \sqrt{\sigma_2^2
+\sigma_3^2+\left( \phi_1^2+\phi_2^2+\phi_3^2\right) -6\sqrt{|q|}
\cos\frac{\theta}{3}} \cdot \nonumber \right. \\ && \cdot \left(
9-\sigma_2^2-\sigma_3^2-\left( \phi_1^2+\phi_2^2+\phi_3^2\right)
+6 \sqrt{|q|} \cos\frac{\theta}{3} \right) \nonumber \\ &+&
(-1)^{\alpha_2} \sqrt{\sigma_2^2 +\sigma_3^2+\left(
\phi_1^2+\phi_2^2+\phi_3^2\right)+3\sqrt{|q|} \left(
\cos\frac{\theta}{3} -\sqrt{3} \sin\frac{\theta}{3}\right)} \cdot
\nonumber \\ && \cdot \left( 9-\sigma_2^2-\sigma_3^2-\left(
\phi_1^2+\phi_2^2+\phi_3^2\right)+3\sqrt{|q|}\left(
-\cos\frac{\theta}{3} +\sqrt{3} \sin\frac{\theta}{3}\right)
\right) \label{32} \\ &+& (-1)^{\alpha_3} \sqrt{\sigma_2^2
+\sigma_3^2+\left(
\phi_1^2+\phi_2^2+\phi_3^2\right)+3\sqrt{|q|}\left(
\cos\frac{\theta}{3}+\sqrt{3}\sin\frac{\theta}{3} \right) }\cdot
\nonumber \\ && \left. \cdot\left( 9-\sigma_2^2-\sigma_3^2-\left(
\phi_1^2+\phi_2^2+\phi_3^2\right)-3\sqrt{|q|} \left(
\cos\frac{\theta}{3} +\sqrt{3}\sin\frac{\theta}{3} \right) \right)
\right] \nonumber
\end{eqnarray}

It is not difficult to identify the singular loci of $W^{(\alpha_1
, \alpha_2 , \alpha_3)}$, and hence of the first-order equations
(\ref{14}), where the super-potential is not continuously
differentiable. As one can easily guess, this happens when
$|q^3+r^2|=0$ and there are two possibilities: (a) if
$|q^3+r^2|=0$ and $r={\rm constant}>0$, i.e. $\theta=0$, the
points of non-differentiability of $W^{(\alpha_1 ,\alpha_2
,\alpha_3)}$ occur at the edge
$\lambda_1=\lambda_2={\bar\sigma}_3^2$ of ${\bf P}_3(\infty)$ or,
in the ellipse :
\begin{equation}
\frac{\phi_1^2}{\sigma_3^2}+\frac{\phi_2^2}{\sigma_3^2-\sigma_2^2}=1
\label{33}
\end{equation}
in ${\Bbb R}^3$. (b)  if, besides $|q^3+r^2|=0$, $r={\rm
constant}<0$, i.e. $\theta=\pi$, the singular curve is
$\lambda_2=\lambda_3={\bar\sigma}_2^2$, or the hyperbola
\begin{equation}
\frac{\phi_1^2}{\sigma_2^2}-\frac{\phi_3^2}{\sigma_3^2-\sigma_2^2}=1
. \label{34}
\end{equation}

We note that the points where the derivatives $\frac{d\phi_a}{dx}$
change sign according the system (\ref{14}) are encoded in the
super-potential itself, not in the metric as in elliptic
coordinates. The $\lambda_3=1$ face, the ellipsoid
\[
\phi_1^2+\frac{\phi_2^2}{{\bar\sigma}_2^2}+\frac{\phi_3^2}{{\bar\sigma}_3^2}=1\qquad
, \label{35}
\]
is a regular zone, however, in Cartesian coordinates.

Finally, it is easy to repeat the calculations in the $N=1$ and
$N=2$ models to obtain the corresponding super-potentials in both
elliptic and Cartesian coordinates. We shall include only the
final results:
\[
W^{(\beta_1,\beta_2)}(\phi_1,\phi_2)=(-1)^{\beta_1} \,
\sqrt{\phi_1^2+\phi_2^2+(-1)^{\beta_2} 2 \sigma_2
\phi_1+\sigma_2^2} \, \left[ \frac{1}{3} \left(
\phi_1^2+\phi_2^2-(-1)^{\beta_2} \sigma_2 \phi_1+\sigma_2^2\right)
-1\right]
\]
$\beta_1,\beta_2=0,1$, are the super-potentials of the $N=2$ model
obtained by reducing to the $\phi_3=0$ plane the previously
analyzed $N=3$ system. An analogous formula occurs in the
$\phi_2=0$ plane, whereas
\[
W^{(\nu)}(\phi_1)=(-1)^{\nu} (\frac{1}{3} \phi_1^3-\phi_1)\qquad ,
\]
$\nu=0,1$, are the super-potentials if the system is reduced to
the $\phi_2=\phi_3=0$ axis.

\end{document}